\documentclass[a4paper,12pt]{article}
\usepackage{amsmath,amssymb}
\usepackage{graphicx,color}
\usepackage{mathrsfs}
\usepackage{cite}
\usepackage{ulem}

  \makeatletter
  \@addtoreset{equation}{section}
  \makeatother
\usepackage[
      colorlinks=true,
      linkcolor=blue,
      urlcolor=blue,
      filecolor=black,
      citecolor=blue,
      pdfstartview=FitV,
      pdftitle={},
        pdfauthor={Ryotaku Suzuki, Shinya Tomizawa},
        pdfsubject={},
        pdfkeywords={},
        pdfpagemode=None,
        bookmarksopen=true,
      ]{hyperref}
\usepackage{caption}

\newcommand{\fr}[1]{\frac{1}{#1}}

\newcommand{\ord}[1]{{\mathcal O}\left(#1\right)}

\newcommand{\cE}{{\mathcal E}}

\newcommand{\cH}{{\mathcal H}}

\newcommand{\cR}{{\mathcal R}}
\newcommand{\cL}{{\mathcal L}}

\newcommand{\nonum}{\nonumber\\ }

\newcommand{\sR}{{\sf R}}

\newcommand{\veps}{ \varepsilon}
\newcommand{\cout}[1]{}

\newcommand{\Li}{{\rm Li}}
\newcommand{\ReLiA}{ {{\rm Dl}_2}}
\newcommand{\ImLiA}{ {{\rm El}_2}}
\newcommand{\ReLiB}{ {{\rm Fl}_2}}
\newcommand{\ImLiB}{ {{\rm Gl}_2}}

\numberwithin{equation}{section}

\begin{document}
%\title{Large $D$ effective theory of black strings in Einstein-Gauss-Bonnet theory}
%\cout{
\begin{titlepage}
\rightline{TTI-MATHPHYS-17}
%\leftline{}{\timestamp}
\vskip 2cm
\vglue 2cm
%\centerline{\LARGE \bf Large $D$ effective theory of black strings}
%\vskip 0.5cm
%\centerline{\LARGE \bf in Einstein-Gauss-Bonnet theory}
\centerline{\LARGE \bf Phase and stability of black strings}
\vskip 0.5cm
\centerline{\LARGE \bf in Einstein-Gauss-Bonnet theory}
\vskip 0.5cm
\centerline{\LARGE \bf at large $D$}

\vskip 1.6 cm
\centerline{\bf Ryotaku Suzuki and Shinya Tomizawa}
\vskip 0.5cm

\centerline{\small Mathematical Physics Laboratory, Toyota Technological Institute}
\centerline{\small 2-12-1 Hisakata, Tempaku-ku, Nagoya 468-8511, Japan}
\smallskip
\vskip 0.5cm
\centerline{\small\tt sryotaku@toyota-ti.ac.jp, tomizawa@toyota-ti.ac.jp}
\vskip 2cm

\centerline{\bf Abstract} \vskip 0.2cm \noindent
\noindent
The phase and stability of black strings in the Einstein-Gauss-Bonnet (EGB) theory are investigated by using the large $D$ effective theory approach. 
The spacetime metric and thermodynamics are derived up to the next-to-leading order (NLO) in the $1/D$ expansion. 
We find that the entropy current defined by the Iyer-Wald formula follows the second law.
As in the Einstein theory, the entropy difference from the total mass produces an entropy functional for the effective theory. Including the NLO correction, we find that for the large Gauss-Bonnet coupling constant $\alpha_{\rm GB}$, the Gregory-Laflamme instability of uniform black strings needs longer wavelength. 
Moreover, we show that the critical dimension, beyond which non-uiform black strings becomes more stable than uniform ones,  increases as $\alpha_{\rm GB}$ becomes large, and approaches to a finite value for $\alpha_{\rm GB}\to \infty$.

\end{titlepage}
\pagestyle{empty}
\small
\addtocontents{toc}{\protect\setcounter{tocdepth}{2}}
{
	\hypersetup{linkcolor=black,linktoc=all}
	\tableofcontents
}
\normalsize

%\newpage
\pagestyle{plain}
\setcounter{page}{1}
%}
%\maketitle
%\abstract{The phase and stability of black strings in the Einstein-Gauss-Bonnet theory are studied via the large $D$ effective theory approach. The effective equation and thermodynamics are derived up to the next-to-leading order in the $1/D$ expansion. We find that the entropy current defined by the Iyer-Wald formula follows the second law, which also provides the entropy functional for the leading order theory. We also show the critical dimension for the stable non-uniform black string increases for the larger Gauss-Bonnet coupling constant.}

\section{Introdcution}

Black holes in dimensions higher than four have been a fascinating subject for the last two decades, motivated by string theory phenomenologies such as the possible production in the particle accelerators with the large extra dimension scenario~\cite{Argyres:1998qn} and 
AdS/CFT correspondence~\cite{Maldacena:1997re}. 
Recent studies reveal that such higher dimensional black holes have rich variety of solutions and dynamics~\cite{Emparan:2008eg}. The simplest and typical example of the rich dynamics in higher dimensions is the black string  constructed as the direct product of a flat spatial direction and a Schwarzschild-Tangerlini black hole~\cite{Tangherlini:1963bw}.
Despite such a simple structure, one can learn from the black string on the generic behavior in higher dimensions such as the Gregory-Laflamme (GL) instability~\cite{Gregory:1993vy,Gregory:1994bj} which leads to the violation of strong cosmic censorship~\cite{Lehner:2010pn} and nonuniquness in the phase diagram~\cite{Gubser:2001ac,Wiseman:2002zc}.
\medskip

The Lovelock theory is the most general theory of gravity including higher curvature corrections whose equations of motion become second order. 
Such a theory arises as the low energy limit by the compactification of $M$ theory  from $11D$ to $5D$. 
In particular, the Einstein-Gauss-Bonnet (EGB) theory with only correction terms of quadratic curvatures  has been
one of major subjects of research for a few decades because of the simplicity.
For $D=4$, this theory coincides with the Einstein gravity since  the higher curvature term  (Gauss-Bonnet terms) vanishes 
but for $D\ge 5$, the action of the theory consists of the Einstein-Hilbelt term and non-vanishing Gauss-Bonnet (GB) terms. 
\medskip

As for black holes in the EGB theory, the first static solution for $D\ge 5$ was found by Boulware and Deser under the assumption of spherical symmetry. 
Finding rotating EGB black hole solution is considered to be a considerably hard problem since the Kerr-Schild formalism does not work in this theory.  
However, the numerical solution of a rotating EGB black hole was obtained in ref.~\cite{Brihaye:2008kh}, and approximate and analytic solutions at the first order in the rotation parameter were found in ref.~\cite{Kim:2007iw}.
Furthermore,  black strings in the EGB theory was studied in refs.~\cite{Kobayashi:2004hq,Suranyi:2008wc,Brihaye:2010me}.

\medskip
Unlike black holes in General Relativity (GR), the thermodynamics of EGB or Lovelock black holes have not been understood well.
The zeroth law for Lovelock black holes was proven assuming the smooth limit to GR~\cite{Ghosh:2020dkk}.
The first law  was shown by Iyer and Wald, who successfully established the general entropy formula~\cite{Wald:1993nt,Iyer:1994ys}.
However, whether the second law with respect to the Iyer-Wald entropy holds even in the Lovelock heory is still an open problem.
By adding appropriate terms, so-called Wall terms, to the Iyer-Wald formula, the second law has been proven to hold in terms of the small variation from stationary solutions~\cite{Wall:2015raa,Bhattacharya:2019qal,Bhattacharyya:2021jhr,Hollands:2022fkn}.
\medskip

The large dimension limit or, large $D$ limit~\cite{Asnin:2007rw,Emparan:2013moa,Emparan:2020inr} is the viable approximation that greatly helps us understand the nonlinear dynamics of black holes in higher dimensions.
At the large $D$ limit, the influence of gravity is localized in a thin layer of $\ord{1/D}$ around the black hole horizon, and the dynamics of the horizon deformation reduces to a simple effective theory on the horizon surface, which we call ``the large $D$ effective theory"~\cite{Emparan:2015hwa,Bhattacharyya:2015dva,Bhattacharyya:2015fdk,Emparan:2015gva}. 
Particularly, the nonlinear dynamics of the black string has been understood very well by the large $D$ effective theory analysis in many aspects such as the Gregory-Laflamme instability, non-uniform branches and critical dimension~\cite{Emparan:2015hwa,Suzuki:2015axa,Emparan:2015gva,Emparan:2018bmi}.
\medskip

Recently, the large $D$ effective theory approach was also applied to the study of EGB black holes, such as the stability analysis of static EGB black holes~\cite{Chen:2017hwm}, dynamics of black strings~\cite{Chen:2017rxa} and black rings~\cite{Chen:2018vbv}. 
Moreover, analytic solutions of rotating black holes with equal angular momenta were also constructed in the $1/D$-expansion without taking the slow rotation limit or small $\alpha_{\rm GB}$ limit~\cite{Suzuki:2022apk}.
% \medskip
The second law of black holes for higher curvature theories was also studied at the large $D$ limit.
The second law, particularly, in the EGB theory proved within the large $D$ membrane paradigm up to $1/D$ and linear order in $\alpha_{\rm GB}$~\cite{Dandekar:2019hyc}.
In ref.~\cite{Saha:2020zho}, appropriate forms of large $D$ effective theories compatible to the second  law were investigated in  general higher curvature theories.
 \medskip
 
In this article, we use the large $D$ effective theory approach to study the horizon dynamics of EGB black string. The effective equations and thermodynamic variables are derived up to the next-to-leading order (NLO) in $1/D$ expansion. Particularly, we show the entropy functional of the effective theory follows the second law, at the nonlinear level.
We show that longer uniform black strings can be stable with the GB correction, that is, $k_{\rm GL}$ becomes longer with larger GB correction.
We also find the GB term admits larger critical dimensions than in GR. 

\medskip
The rest of the article is organized as follows. 
First, we derive the metric functions using $1/D$-expansion in section~\ref{sec:metric}.
In section~\ref{sec:eft}, the large $D$ effective theory of the EGB black string are studied.
Solving the effective equation, we study the phase and stability of uniform and nonuniform black strings in section~\ref{sec:staticphase}.
Finally, we summarize our result in section~\ref{sec:sum}.

\section{Metric solution}\label{sec:metric}
We consider the EGB theory, whose action is given by
\begin{align}
 S = \fr{16\pi G} \int \sqrt{-g}\left(R  + \alpha_{\rm GB} \cL_{\rm GB}\right)d^Dx,
\end{align}
where the GB correction term $\cL_{\rm GB}$ is written as
\begin{align}
\cL_{\rm GB} = R^2-4R_{\alpha\beta} R^{\alpha\beta} + R_{\alpha\beta\gamma\delta}R^{\alpha\beta\gamma\delta}.
\end{align}
This action leads to the EGB equation
\begin{align}
 R_{\mu\nu} - \fr{2}Rg_{\mu\nu}+ \alpha_{\rm GB} H_{\mu\nu} =0\label{eq:EGB-eom},
\end{align}
where
\begin{align}
 H_{\mu\nu} = 2RR_{\mu\nu}-4 R_{\mu\alpha}R^\alpha{}_\nu-4R_{\mu\alpha\nu\beta}R^{\alpha\beta} + 2 R_{\mu\alpha\beta\gamma}R_\nu{}^{\alpha\beta\gamma}-\fr{2} \cL_{\rm GB} g_{\mu\nu}.
\end{align}
The large $D$ limit can lead to different limits depending on the scale assumption of the coupling constant $\alpha_{\rm GB}$ at the limit. We choose the scaling so that the Einstein-Hilbert term and the GB term remain  comparable.
At large $D$, the radial gradient on the horizon is roughly estimated as $\partial_r = \ord{D}$~\cite{Emparan:2013moa}, and hence, each term in the action  is estimated as $R \sim g^{tt}\partial_r^2 g_{tt} = \ord{D^2}$ and ${\cal L}_{\rm GB}\sim R^2 = \ord{D^4}$. 
Then, we must assume the $\alpha_{\rm GB}$ scale as $\ord{D^{-2}}$. It turns out that the other cases are easily obtained as the parameter limits of $\alpha_{\rm GB} \to 0$ or $\alpha_{\rm GB} \to \infty$, after the EGB equations are solved in the $1/D$ expansion. 
Due to the lengthy equations in the $1/D$ expansion, we do not write them here. 
See Appendix A for the detail. 

\subsection{Setup}
To obtain the $D=n+4$ dynamical black string, we assume the following ansatz
in the Eddington-Finkelstein coordinates
\begin{align}
ds^2 = - A dt^2 + 2 U dt dr - \fr{n}C dt dz + \fr{n} G dz^2 + r^2 d\Omega_{n+1}^2,\label{eq:setup-ansatz}
\end{align}
where $d\Omega_{n+1}^2$ denotes the line element of $S^{n+1}$.
 For convenience, we use $n$ as the large parameter instead of $D$ in the following analysis.
As the boundary condition, we impose that the spacetime is asymptotically flat in the direction transverse to the string
\begin{align}
 A(r\to\infty)  = 1 ,\quad U(r\to\infty) = 1,\quad C(r\to\infty) = 0,\quad G(r\to\infty) = 1,
\end{align}
and regular on the horizon. In the ansatz~(\ref{eq:setup-ansatz}), 
the string direction $z$ compactified by $z\sim z+L$ is rescaled by $1/\sqrt{n}$ in advance, to capture the Gregory-Laflamme instability of the wavenumber of $\ord{\sqrt{n}}$~\cite{Asnin:2007rw,Emparan:2013moa}.

To expand the metric in $1/n$, we assume that the horizon is placed at $r=r_0+\ord{1/n}$, and introduce the near-horizon coordinate by
\begin{align}
 \sR := (r/r_0)^n.
\end{align}
With the near-horizon coordinate, the metric functions are expanded as
\begin{align}
& A = \sum_{i=0} \frac{A_i(t,{\sf R},z)}{n^i},\quad C = \sum_{i=0} \frac{C_i(t,{\sf R},z)}{n^i} ,\nonum
 & G = 1+\fr{n} \sum_{i=0}\frac{G_i(t,{\sf R},z)}{n^i},\quad U = 1 + \fr{n}\sum_{i=0} \frac{U_i(t,{\sf R},z)}{n^i}.\label{eq:metricexp}
\end{align}
We also define the rescaled coupling constant which remains $\ord{n^0}$ by
\begin{align}
\alpha := \frac{n^2 \alpha_{\rm GB}}{r_0^2}.
\end{align}
In the following, we set the radius scale as $r_0=1$ by changing the overall scale.

\subsection{Leading order analysis}
At the leading order, $A_0$ is written by solving eq.~(\ref{eq:LOeq-A0}), as
\begin{align}
 A_0 = 1+\fr{2\alpha}-\fr{2\alpha}\sqrt{1+\frac{4\alpha(\alpha+1)m(t,z)}{\sR}},\label{eq:LOsolA}
\end{align}
where we consider only the $(-)$-branch that allows the GR limit at $\alpha \to 0$. $m(t,z)$ is an integration function with respect to the $\sR$-integral, which determines the horizon position ${\sf R}=m(t,z)$ as well as the mass density at ${\sf R} \to \infty$. From eq.~(\ref{eq:LOsolA}) and eq.~(\ref{eq:LOeq-C0}), $C_0$ is also obtained 
\begin{align}
 C_0 = \frac{p(t,z)}{2\alpha m(t,z)}\left( \sqrt{1+\frac{4\alpha(\alpha+1)m(t,z)}{\sf R}}-1\right),
\end{align}
with the other integration function $p(t,z)$ that arouse the non-uniformity and horizon velocity along the string direction. For $G_0$ and $U_0$, it is convenient to switch to  the auxiliary variable
\begin{align}
 X := \sqrt{1+\frac{4\alpha(\alpha+1)m}{\sf R}},\label{eq:def-X}
\end{align}
which takes the value between $X=1\ (\sR=\infty)$ and $X=1+2\alpha\ (\sR=m)$.
$A_0$ and $C_0$ are rewritten in terms of $X$ as
\begin{align}
 A_0 = \frac{1+2\alpha-X}{2\alpha},\quad C_0 = \frac{p}{2\alpha m}(X-1).
\end{align}
Then, $G_0$ and $U_0$ are obtained by solving eqs.~(\ref{eq:LOeq-G0}) and (\ref{eq:LOeq-U0}), respectively, as
\begin{align}
 G_0& = \left(\frac{\log \left(X^2+1\right)}{2\alpha+1}-2\log(X+1)+2\arctan X\right) \left(\frac{ m\partial_z p-p\partial_z m + m^2   }{(\alpha +1) m ^2}\right)\nonum
   &-\frac{((2  \alpha+1)\pi -2 (4 \alpha+1)  \log 2 ) (  m \partial_z p - p\partial_z m  + m^2) }{2 (\alpha +1) (2 \alpha +1) m ^2}+\frac{ p ^2}{2 \alpha  m ^2}(X-1),
\end{align}
and
\begin{align}
U_0 &=
 \frac{p  \left(\left(2 \alpha ^2+3 \alpha +1\right) p -4 \alpha ^2
    \partial_z m  \right)+4 \alpha ^2 m   \partial_z p  +4 \alpha ^2 m ^2}{4 \alpha  (\alpha +1) (2 \alpha +1) m ^2}\nonum
    &-\frac{((2 \alpha+1)  X \partial_z m -1) \left(- \partial_z m   p +m   \partial_z p  +m ^2\right)}{(\alpha +1) (2 \alpha +1) \left(X^2+1\right) m ^2}-\frac{X p ^2}{4 \alpha  m ^2}.
\end{align}

\paragraph{Constraints}
Plugging the leading order solutions to eqs.~(\ref{eq:constT}) and (\ref{eq:constZ}), we obtain the leading order effective equation, which is first shown in ref.~\cite{Chen:2017rxa},
\begin{align}
& \partial_t m - \partial_z^2 m = - \partial_z p,\\
& \partial_t p - \partial_z^2 p = \partial_z \left( m-\frac{p^2}{m} +
\frac{2\alpha}{(\alpha+1)(2\alpha+1)}\left(\frac{p \partial_z m-m \partial_z p- m^2}{m}\right)\right).\label{eq:LOeq}
\end{align}

\subsection{Next-to-Leading order solution}
At the higher order, after imposing the same boundary condition as above, eq.~(\ref{eq:nloeq}) provides the extra undetermined functions in $A_i$ and $C_i$ that corresponds to the shift of $\ord{n^{-i}}$ in $m$ and $p$. We choose these functions so that $A_i(\sR=m) =0$ and $C_i(\sR=m)=0$ for $i\geq 1$. This choice sets the event horizon at $\sR=m$ for the static solution at each order. 

$A_1$ is obtained by solving eq.(\ref{eq:nloeq-A}) at the next-to-leading order (NLO), 
\begin{align}
& A_1 =\frac{(X^2-1)\left(- p\partial_z m   +m 
    \partial_z p  +m ^2\right)}{4 \alpha (\alpha +1) X
   m ^2}\left(\frac{\log(X^2+1)}{2(2\alpha+1)}+\arctan X\right)\nonum
&    +\frac{(X-1) \log (X+1)
   \left((X+1)  p\partial_z m    +\alpha  (X+1) m   \partial_z p  +(-2 \alpha +2 \alpha  X+X-3)
   m ^2\right)}{4 \alpha  (\alpha +1) X m ^2}\nonum
   &+\frac{(X-1) \log (X-1) \left(2 (X-1)
   m +(X+1)  \partial_z p  \right)}{4 \alpha  X m }\nonum
 &+\frac{(X-1) \left(p^2 -2 p \partial_z m +2 m   \partial_z p  +m ^2 (2 \log (4\alpha(\alpha+1) m)+1)\right)}{2 \alpha  X m ^2}\nonum
 & + \frac{(1-X^2)(2(2\alpha+1)(2m^2\log m+p^2)+c_{1} m^2+c_{2} p \partial_z m + c_{3} m \partial_z p) }{8X\alpha(1+\alpha)(2\alpha+1) m^2},
\end{align}
where the coefficients $c_1,c_2,c_3$ are given by
\begin{align}
 & c_{1} = 2(2\alpha+1)(1+\arctan(2\alpha+1))+(16\alpha^2+20\alpha+6)\log 2+4(2\alpha+1)(\alpha+1)\log \alpha\nonum
 & \qquad+ 2(2\alpha+1)^2 \log(\alpha+1)+\log(1+(2\alpha+1)^2),\\
 & c_2 = 2 (2\alpha+1) (-2-\arctan(2\alpha+1)+\log 2+\log(\alpha+1))-\log(1+(2\alpha+1)^2),\\
 & c_3 = 2 (2\alpha+1) (2+\arctan(2\alpha+1))+2(2\alpha+1)^2\log 2+ 2(2\alpha+1)(\alpha+1)\log \alpha
 \nonum
 &\qquad+ 2\alpha(2\alpha+1)\log(\alpha+1)+\log(1+(2\alpha+1)^2).
\end{align}
Due to the considerably lengthy form, we do not explicitly write other components.
Instead, we roughly explain the form of the solutions.
$C_1$ and $U_1$ are much lengthy and are expressed with the combination of $\log(X\pm1),\log(X^2+1),\arctan X$ and rational functions of $X$, as is $A_1$. $H_1$ is written in the combination of the polylogarisms
\begin{align}
\Li_2\left(\frac{1-X}{2}\right),\quad \Li_2\left(\frac{1+2\alpha-X}{2(1+\alpha)}\right),\quad\Li_2\left(\frac{1+2\alpha-X}{2\alpha}\right),
\end{align}
Clausen's functions
 \begin{align}
 {\rm Cl}_2(2 \arctan X),\quad  {\rm Cl}_2(2 \arctan X+\pi/2),\quad 
  {\rm Cl}_2(2 \arctan X+\pi),\quad   {\rm Cl}_2(2 \arctan X+3\pi/2),
 \end{align}
 and some complex polylogarisms given in Appendix~\ref{sec:funcs}
 \begin{align}
\ReLiA(X),\quad \ReLiA(-X),\quad \ReLiB(X),\quad  \ImLiB(X),
 \end{align}
 in addition to $\log(X\pm1),\log(X^2+1),\arctan X$ and rational functions of $X$.

\section{Large $D$ effective theory}\label{sec:eft}
Substituting the NLO solutions in the previous section into eqs.~(\ref{eq:constT}) and (\ref{eq:constZ}),
we obtain the effective equation up to NLO correcting eq.~(\ref{eq:LOeq}).
To keep the readability, we do not write the explicit form of the lengthy correction, but instead express it in terms of 
the conservation law of the quasi-local stress energy tensor
\begin{align}
 \partial_t  T^{tt} + \partial_z T^{tz} = 0,\quad  \partial_t  T^{tz} + \partial_z T^{zz} = 0,\label{eq:NLOeq-conserve}
\end{align}
where $T^{\mu\nu}$ is shown in Appendix~\ref{sec:brownyork}.
The effective equation is invariant under the scaling law that reflects the freedom to choose the overall scale of the geometry
\begin{align}
 m \to C m,\quad  p \to C p, \quad \partial_t \to C^{-1/n} \partial_t, \quad\partial_z \to C^{-1/n} \partial_z,\quad \alpha \to  C^{2/n} \alpha, \label{eq:scaling-law}
\end{align}
where the scaling law for $\alpha$ reflects the dimension of $\alpha_{\rm GB}$.

\subsection{The second law}
In the previous work~\cite{Chen:2017rxa}, the metric solution is derived only up to the leading order, where the entropy is found trivially proportional to the ADM mass as in the GR case~\cite{Emparan:2013moa}.
We show that the NLO correction to the entropy is crucial to admits the second law of black holes in the effective theory.

\subsubsection{Entropy current}
We construct the entropy current from the dynamical effective theory in terms of the local event horizon~\cite{Bhattacharyya:2008xc}, in which the area factor is replaced with the Iyer-Wald formula~\cite{Wald:1993nt,Iyer:1994ys}\footnote{It has been shown that we do not need the modification to the entropy for the second law in the EGB theory~\cite{Wall:2015raa}.}.
The local event horizon $H$ is defined as the null surface $r - r_H(t,z)=0$ which satisfies
\begin{align}
A-2 U \partial_t r_H + \fr{n}G^{-1}(C-n U \partial_z r_H)(C- n U \partial_z r_H)=0.
\end{align}
The spatial metric of the local event horizon $H$ is given by
\begin{align}
  ds^2_H = \fr{n}G_H (dz-v dt)^2+r_H^2 d\Omega_{n+1}^2,\label{eq:metricH}
\end{align}
where we define the velocity field $v$ as
\begin{align}
v = C_H-U_H \partial_z r_H.
\end{align}
With the near-horizon coordinate $\sR_H := r_H^n$, the large $D$ limit leads to 
\begin{align}
A\bigr|_{\sR=\sR_H} +\ord{1/n}=0 \quad \Rightarrow  \quad \sR_H  = m+\ord{1/n},
\end{align}
and
\begin{align}
 v = (p-\partial_z m)/m + \ord{1/n}.\label{eq:def-vlo}
\end{align}
With the NLO solution, these are corrected as
\begin{align}
\sR_H = m-\frac{(2 \alpha +1) \left((p- \partial_z m)^2-2 m \partial_t m\right)}{n(\alpha +1) m}
\end{align}
and
\begin{align}
v& =  \frac{p-\partial_z m}{m}+\fr{n}\left[\frac{2(2 \alpha +1) \partial_z m ( \partial_t m+\partial_z^2 m)}{(\alpha +1) m ^2}
-\frac{\left(16   \alpha ^4+34 \alpha ^3+28 \alpha ^2+11 \alpha +2\right) \partial_z m \partial_z p }{(\alpha +1) (2 \alpha +1) \left(2 \alpha ^2+2 \alpha +1\right)   m ^2}
\right.\nonum
&   +\frac{\alpha(1 -2 \alpha ^2) \partial_z m}{(\alpha +1) (2 \alpha +1) \left(2 \alpha ^2+2 \alpha
   +1\right) m  } -\frac{2 (2 \alpha +1)(m p(\partial_z^2 m-\partial_z p)+(\partial_z m)^3+m^2 \partial_t \partial_z m)  }{(\alpha +1) m ^3}\nonum
& \left.  +\frac{\left(36 \alpha ^4+76 \alpha ^3+66 \alpha ^2+28 \alpha +5\right) (\partial_z m) ^2 p }{(\alpha +1) (2 \alpha +1)   \left(2 \alpha ^2+2 \alpha +1\right) m ^3}+\frac{p ^3}{m ^3}-\frac{(11 \alpha +7) p ^2 \partial_z m}{2 (\alpha +1) m ^3}-\frac{2 \partial_t m   p }{m ^2}-\frac{\partial_z m  \log m}{m }  \right].
\label{eq:def-vnlo}
\end{align}
Then, the entropy current is given by
\begin{align}
 {\bf J}_S^\mu \partial_\mu  := \frac{\Omega_{n+1}}{4G\sqrt{n}} (\rho_S \partial_t + j_S\partial_z)
\end{align}
where
\begin{align}
 \rho_S :=  (1+ 2 \alpha R_H)\sqrt{G_H}r_H^{n+1},\quad j_S :=  \rho_S v
\end{align}
and $R_H$ is the scalar curvature of the horizon surface~(\ref{eq:metricH}).

We find that, up to NLO, the entropy density is given by
\begin{align}
\rho_S& = (2 \alpha +1) m \nonum
&-\frac{1}{4
   (\alpha +1) n }\left[  \partial_z p  \left((2 \alpha+1)(\pi -4 \arctan(2 \alpha +1)
  +4\log(\alpha+1))
  -2 \log(2\alpha^2+2\alpha+1)   \right)\right. \nonum
&\left.   -8 (2 \alpha +1)^2 \partial_t m 
   +16 \alpha  (\alpha +1)   \partial_z^2 m + 4 \left(2 \alpha ^2+\alpha -1\right) m\log  m \right.\nonum
&\left.   
   -\frac{\partial_z m  p}{m} \left(8 (2\alpha+1)^2+(2 \alpha +1)(\pi 
   -4 \arctan (2 \alpha +1)+4 \log(\alpha+1))
   -2 \log(2\alpha^2+2\alpha+1) \right)\right. \nonum
&\left.    +m \left(
(2 \alpha+1)(\pi  -4 \arctan(2 \alpha +1)+4\log(\alpha+1))-2 \log(2\alpha^2+2\alpha+1)-8 \alpha(\alpha+1)\right)\right.\nonum
   &\left. +4 \left(2 \alpha ^2+2
   \alpha +1\right) \frac{\partial_z m ^2}{m}+2 (2\alpha+1)(3\alpha+1) \frac{p^2}{m}
  \right].\label{eq:ent-dense}
\end{align}
With the entropy density, we define the dynamical entropy for the effective theory up to NLO
\begin{align}
 S := \int_0^L \rho_S dz.
\end{align}
Using the effective equation~(\ref{eq:NLOeq-conserve}), one can verify\footnote{We need to add some total derivative terms to the flux part to satisfy the local second law. Nevertheless, this does not affect the second law in the integrated form.}
\begin{align}
  \partial_t \rho_S + \partial_z \left(j_S +1/n( \dots)\right)= \frac{2(1+\alpha+2\alpha^2)}{(1+\alpha)n} m (\partial_z v)^2 \geq 0,
\end{align}
where $p$ is replaced by $v$ using eq.~(\ref{eq:def-vlo}).
This is consistent with the result by the large $D$ approach in~\cite{Dandekar:2019hyc} up to the linear order in $\alpha$.

The second law in the large $D$ effective theory follows from this inequality
\begin{align}
 \partial_t S =\frac{2(1+\alpha+2\alpha^2)}{(1+\alpha)n} \int_0^L m (\partial_z v)^2 dz \geq 0.\label{eq:2ndlaw}
\end{align}
The entropy-production rate only differs from that in GR by the factor depending on the GB coupling constant
\begin{align}
 \frac{\partial_t S}{S} = \frac{2\alpha^2+\alpha+1}{(2\alpha+1)(\alpha+1)}\frac{\displaystyle\int_0^L 2m (\partial_z v)^2 dz}{\displaystyle n\int_0^L m  dz}.
\end{align}
For a given configuration $m(t,z)$ and $v(t,z)$, the entropy-production rate is minimized at $\alpha = 1/\sqrt{2}$ and goes back to the GR rate at $\alpha \to \infty$. 
This is consistent with the fact that the time evolution from a uniform to a non-uniform black string takes a largest duration at $\alpha = 0.708 \approx 1/\sqrt{2}$ in ref.~\cite{Chen:2017rxa}.

\subsubsection{Entropy functional for effective theory}
The entropy density~(\ref{eq:ent-dense}) is proportional to the mass density~(\ref{eq:defTtt}) at the leading order. Therefore, as in the GR case~\cite{Andrade:2020ilm}, we can construct a monotonic functional for the leading order theory 
by taking the difference between the entropy and mass so that the leading order terms cancel,
\begin{align}
&n \left[\rho_S - \frac{2\alpha+1}{\alpha+1}\left(1+\frac{2\alpha^2-1}{(\alpha+1)(2\alpha+1)}\right)T^{tt}\right] \nonum
& \quad = \frac{1+\alpha+2\alpha^2}{1+\alpha}  \left(- \frac{(\partial_z m)^2}{2m}- \frac{(2\alpha+1)(1+\alpha)}{2(1+\alpha+2\alpha^2)}mv^2 + m \log m\right)
+\partial_z (\dots),
\end{align}
where $v$ is given by eq.~(\ref{eq:def-vlo}). The coefficient of $T^{tt}$ is chosen so that the right hand side in the above expression does not include the term ${\rm const.} \times m$.\footnote{The term $C m$ only produces $C M$ in eq.~(\ref{eq:def-S1}) which remains constant during the time evolution. } Therefore, we define the entropy functional by
\begin{align}
S_1 =  \frac{1+\alpha+2\alpha^2}{1+\alpha} \int_0^L \left(- \frac{(\partial_z m)^2}{2m}- \frac{(2\alpha+1)(1+\alpha)}{2(1+\alpha+2\alpha^2)}mv^2 + m \log m\right)dz.\label{eq:def-S1}
\end{align}
The second law for the entire entropy~(\ref{eq:2ndlaw}) guarantees the monotonicity of this functional,
\begin{align}
\partial_t  S_1 = \frac{2(1+\alpha+2\alpha^2)}{1+\alpha}   \int_0^L m (\partial_z v)^2 dz \geq 0.
\end{align}
One can also confirm the monotonicity using only the leading order effective equation~(\ref{eq:LOeq}).

\subsection{Static equation}
In the static case, eq.~(\ref{eq:NLOeq-conserve}) reduces to a simple master equation.
The static solution is given by assuming $T^{tz}=0$ or $v=0$ which leads to
\begin{align}
& p(z) = m'(z) +\fr{n}\left[\left(\frac{\alpha  \left(2
   \alpha ^2-1\right) }{4 \alpha ^4+10 \alpha
   ^3+10 \alpha ^2+5 \alpha +1}+\log m(z)\right) m'(z)\right.\nonum
&  \left.-\frac{\left(4 \alpha
   ^4+14 \alpha ^3+10 \alpha ^2+3 \alpha +1\right)
   m'(z)^3}{2 (\alpha +1) (2 \alpha +1) \left(2 \alpha
   ^2+2 \alpha +1\right) m(z)^2}+\frac{\alpha  \left(2 \alpha ^2-1\right) m'(z)
   m''(z)}{(\alpha +1) (2 \alpha +1) \left(2 \alpha^2+2 \alpha +1\right) m(z)}\right].
   \label{eq:static-p}
\end{align}
With this condition, eq.~(\ref{eq:NLOeq-conserve}) reduces to a third order ODE for $m(z)$. By introducing $\cR(z)=\log m(z)$, it can be integrated to the second order ODE
\begin{align}
\cR''(z)+\frac{1}{2} \cR'(z)^2+\cR(z)
= \fr{n}\left(c_{01} \cR(z)+ c_{02} \cR(z)^2+(c_{20}+c_{21}\cR(z))\cR'(z)^2+\fr{4} c_{21}  \cR'(z)^4\right),\label{eq:static-eq}
\end{align}
where the integration constant is fixed so that $\cR=0$ gives the uniform solution by using the scaling law~(\ref{eq:scaling-law}). The coefficients are given by
\begin{subequations}\label{eq:static-eq-cfs}
\begin{align}
& c_{01} = \frac{20 \alpha ^3+2 \pi  \alpha ^2+34 \alpha ^2+2 \pi
    \alpha +12 \alpha +\pi +2}{2(\alpha+1)(2\alpha^2+\alpha+1)}+\frac{2 \log (\alpha +1)}{(\alpha
   +1) \left(2 \alpha ^2+\alpha +1\right)}\nonum
& \quad  -\frac{2   \left(2 \alpha ^2+2 \alpha +1\right) \arctan(2
   \alpha +1)}{(\alpha +1) \left(2 \alpha ^2+\alpha
   +1\right)}-\frac{\left(2 \alpha ^2+2 \alpha
   +1\right) \log(2\alpha^2+2\alpha+1)}{(\alpha+1)(2\alpha+1)(2\alpha^2+\alpha+1)},\label{eq:static-eq-cfs01}\\
& c_{02} = -\frac{12 \alpha ^3+22 \alpha ^2+\pi  \left(4 \alpha
   ^2+3 \alpha +2\right)+4 \alpha -2}{4 (\alpha +1)
   \left(2 \alpha ^2+\alpha +1\right)}-\frac{\left(2
   \alpha ^2+\alpha +2\right) \log (\alpha
   +1)}{(\alpha +1) \left(2 \alpha ^2+\alpha
   +1\right)}\nonum
  &\quad+\frac{\left(4 \alpha ^2+3 \alpha
   +2\right) \arctan(2 \alpha +1)}{(\alpha +1)
   \left(2 \alpha ^2+\alpha +1\right)}+\frac{\left(4
   \alpha ^2+3 \alpha +2\right) \log(2\alpha^2+2\alpha+1)}{8 \alpha
   ^4+16 \alpha ^3+14 \alpha ^2+8 \alpha +2},   \\
& c_{20}=\frac{12 \alpha ^3+2 \pi  \alpha ^2+22 \alpha ^2+2 \pi
    \alpha +4 \alpha +\pi -2}{4 (\alpha +1) \left(2
   \alpha ^2+\alpha +1\right)}+\frac{\log (\alpha
   +1)}{2 \alpha ^3+3 \alpha ^2+2 \alpha
   +1},\nonum
&\quad   -\frac{\left(2 \alpha ^2+2 \alpha +1\right) \tan
   ^{-1}(2 \alpha +1)}{(\alpha +1) \left(2 \alpha
   ^2+\alpha +1\right)}-\frac{\left(2 \alpha ^2+2
   \alpha +1\right) \log(2\alpha^2+2\alpha+1)}{8 \alpha ^4+16 \alpha
   ^3+14 \alpha ^2+8 \alpha +2},\\
& c_{21}=-\frac{36 \alpha ^3+58 \alpha ^2+\pi  \left(4 \alpha
   ^2+3 \alpha +2\right)+28 \alpha +10}{4 (\alpha +1)
   \left(2 \alpha ^2+\alpha +1\right)}-\frac{\left(2
   \alpha ^2+\alpha +2\right) \log (\alpha
   +1)}{(\alpha +1) \left(2 \alpha ^2+\alpha
   +1\right)}\nonum
   &\quad+\frac{\left(4 \alpha ^2+3 \alpha
   +2\right) \arctan (2 \alpha +1)}{(\alpha +1)
   \left(2 \alpha ^2+\alpha +1\right)}+\frac{\left(4
   \alpha ^2+3 \alpha +2\right) \log(2\alpha^2+2\alpha+1)}{8 \alpha
   ^4+16 \alpha ^3+14 \alpha ^2+8 \alpha +2}.
\end{align}
\end{subequations}

\paragraph{Non-uniformity}
To parametrize the non-uniform phase,
it is convenient to define a scale-independent measure of the static deformation.
Here we introduce the non-uniformity parameter as\footnote{We adopt the convention convenient for the large $D$ analysis, which is different from the earlier works~\cite{Gubser:2001ac}.}
\begin{align}
 \lambda := \fr{2}\left( \frac{e^{\cR_{\rm max}}}{e^{\cR_{\rm min}}}-1\right).
\label{eq:def-lambda}
\end{align}

\subsubsection{Thermodynamics}
Once eq.~(\ref{eq:static-eq}) is solved, the thermodynamics phase is calculated as
\begin{align}
 &\textsc{Mass} = \frac{n\Omega_{n+1}}{16 \pi G \sqrt{n}} \int_0^L T^{tt} dz =:   \frac{n\Omega_{n+1}}{16 \pi G \sqrt{n}}M,\\
& \textsc{Entropy} =: \frac{\Omega_{n+1}}{4 G \sqrt{n}} S,\\
 & \textsc{Temperature} =: \frac{n}{4\pi} T_H,\\
  & \textsc{Tension}= -\frac{\Omega_{n+1}}{16\pi G\sqrt{n}} \frac{T^{zz}}{n} =: \frac{\Omega_{n+1}}{16\pi G \sqrt{n}} {\cal T},
\end{align}
where
\begin{align}
M& = (\alpha+1)\int_0^L dz  e^{\cR} \left[1+\fr{n}\left(\frac{4\alpha+3-(\alpha-1)\cR}{\alpha+1}-\left(\frac{3 \alpha+2}{\alpha+1} +\cR\right)   \cR'^2 \right.\right.\nonum
 &\left.\left.+\left(\frac{\arctan(2 \alpha
   +1)-\pi/4-2(2 \alpha+1)-\log(\alpha+1)}{\alpha+1}+\frac{\log(2\alpha^2+2\alpha+1)}{2(2 \alpha +1)(\alpha+1)}- \cR(z)\right)(\cR''+1)\right)\right],\\
  S& = (2\alpha+1)\int_0^L dz  e^{\cR} \left[1+\fr{n}\left( \frac{6 \alpha+(1-2\alpha) \cR}{2\alpha+1} -\fr{2} \frac{2\alpha-1}{2\alpha+1} (\cR')^2\right.\right.\nonum
 &\left.\left.-\left(\frac{\arctan(2 \alpha
   +1)-\pi/4-\log(\alpha+1)}{\alpha+1}-\frac{4\alpha}{2\alpha+1}+\frac{\log(2\alpha^2+2\alpha+1)}{2(2 \alpha +1)(\alpha+1)}\right)(\cR''+1)\right)\right], 
\end{align}
\begin{align}
 T_H =\frac{\alpha+1}{2\alpha+1}\left[1 - \fr{n}\left(\frac{\alpha  (4 \alpha +1)}{2 (2 \alpha +1)(\alpha+1)}+\frac{\left(2 \alpha ^2+\alpha +1\right)
   }{ (2 \alpha +1)(\alpha+1)}\left( \cR''(z)+\fr{2}\cR'(z)^2+\cR(z)\right)\right)\right],\label{eq:temp-st}
\end{align}
and
\begin{align}
{ \cal  T}& = \frac{e^\cR}{n}\left[\frac{2\alpha^2+\alpha+1}{2\alpha+1} (1+\cR'')\right.\nonum
&\left.+\fr{n}\left(t_0+t_1\cR+t_2(\cR')^2+(t_3+t_4\cR)\cR''+t_5(\cR'')^2+t_6(\cR')^2\cR''+t_7\cR'\cR^{(3)}+t_8\cR^{(4)}\right) \right].
\end{align}
The coefficients for ${\cal T}$ is given by
\begin{align}
& t_{0} =\frac{2 \alpha ^2+\alpha +1}{(2\alpha+1)(\alpha+1)}  \left(\frac{ \log \left(2 \alpha ^2+2 \alpha +1\right)}{2   (2 \alpha +1)}
   +\arctan(2 \alpha +1)-\log  (\alpha +1)-\frac{\pi}{4}\right) \nonum
   &\hspace{7cm}+\frac{\alpha  \left(4 \alpha ^3-4 \alpha ^2-11 \alpha -3\right)}{(\alpha +1) (2  \alpha +1)^2},\nonum
& t_{1} =-\frac{2 \alpha  \left(4 \alpha ^2+4 \alpha -1\right)}{(2 \alpha +1)^2},\quad t_2 = \frac{4\alpha+1}{(2\alpha+1)^2},
\quad t_{4} = \frac{2(4\alpha+1)}{(2\alpha+1)^2},\quad t_6 = - \frac{8\alpha^3+8\alpha^2+2\alpha+1}{(2\alpha+1)^2},\nonum
&t_3=\frac{1+9 \alpha +20 \alpha ^2+32 \alpha ^3+16 \alpha ^4}{ (1+\alpha )
   (1+2 \alpha )^2}+\frac{\pi}{2}-2 \arctan(1+2 \alpha )\nonum
&\hspace{5cm}   +\frac{(2-4 \alpha ) \log (1+\alpha )-\log   \left(1+2 \alpha +2 \alpha ^2\right)}{1+2 \alpha },\nonum
& t_5 = \frac{1+12 \alpha +31 \alpha ^2+36 \alpha ^3+12 \alpha ^4}{(1+\alpha ) (1+2 \alpha )^2}+\frac{\left(3-\alpha   -2 \alpha ^2\right) \log (1+\alpha )}{(1+\alpha ) (1+2 \alpha )}\nonum
&\hspace{1cm}+\frac{\left(3+7 \alpha +6 \alpha
   ^2\right) \left((1+2\alpha)\pi -4 (1+2 \alpha ) \arctan(1+2 \alpha )-2 \log \left(1+2 \alpha +2
   \alpha ^2\right)\right)}{4 (1+\alpha ) (1+2 \alpha )^2},\nonum
&t_7=\frac{2 +8 \alpha  +14 \alpha ^2+4 \alpha ^3- \left(1+2 \alpha +2 \alpha
   ^2\right) (4\arctan(1+2 \alpha )-\pi)+4 \log (1+\alpha )}{2+6 \alpha +4 \alpha ^2}\nonum
&\hspace{5cm}   -\frac{\left(1+2 \alpha +2
   \alpha ^2\right) \log \left(1+2 \alpha +2 \alpha ^2\right)}{(1+\alpha ) (1+2 \alpha )^2},\nonum
& t_8 =\frac{2 \alpha }{1+ \alpha }+\frac{\pi}{4}-\arctan(1+2 \alpha )+\frac{(1-2 \alpha ) \log (1+\alpha
   )}{1+2 \alpha }-\frac{\log \left(1+2 \alpha +2 \alpha ^2\right)}{2+4 \alpha }.
\end{align}
The Hawking temperature is calculated in terms of the surface gravity
\begin{align}
\textsc{Temperature} = \frac{n}{4\pi} T_H = \frac{\kappa}{2\pi} = \left. \frac{1}{2\pi}\frac{\partial_r A}{2 U}\right|_{r=r_H}.
%T_H = \frac{2\kappa}{n} =\left. \frac{2}{n}\frac{\partial_r A}{2 U}\right|_{r=r_H}.
\end{align}
Up to NLO, the scaling law~(\ref{eq:scaling-law}) acts on these variables as
\begin{align}
 M \to C^\frac{n+1}{n} M ,\quad S \to C^\frac{n+2}{n} S ,\quad {\cal T} \to C{\cal T},\quad T_H \to C^{-1/n} T_H.
\end{align}
In eq.~(\ref{eq:static-eq}), we have already fixed the scaling so that the uniform solution is uniquely given by $\cR=0$ up to NLO. This scaling choice corresponds to fixing the temperature (\ref{eq:temp-st}) to
\begin{align}
T_H =\frac{\alpha+1}{2\alpha+1}\left[1 - \frac{\alpha  (4 \alpha +1)}{2 n(2 \alpha +1)(\alpha+1)}\right].\label{eq:temp-st-fix}
\end{align}

\paragraph{First law}
With the Iyer-Wald entropy, the static solution satisfies the first law
\begin{align}
dM = T_H dS+{\cal T} dL.
\end{align}
If we recover the radius scale $r_0$, the thermodynamic variables can be written as
\begin{align}
  M = r_0^{n+1} \hat{M}(\alpha,\lambda),\quad
    S = r_0^{n+2} \hat{M}(\alpha,\lambda),\quad
    {\cal T} =   r_0^{n} \hat{{\cal T}}(\alpha,\lambda),\quad
      T_H = r_0^{-1} \hat{T_H}(\alpha,\lambda),
\end{align}
where the quantities with hat are scale invariant parts.
By considering $r_0$ and $\lambda$ as independent parameters and recalling the definition $\alpha=n^2\alpha_{\rm GB}/r_0^2$, the first law reduces to
\begin{align}
&dM - T_H dS-{\cal T} dL\nonum
&\quad =\bigl[(n+1)M - (n+2)T_H S-{\cal T} L-2\alpha \left(\partial_{\alpha} M
-T_H \partial_{\alpha} S- {\cal T} \partial_{\alpha} L \right)\bigr] \frac{dr_0}{r_0}\nonum
& \quad\quad+ \left(\partial_\lambda M - T_H \partial_\lambda S - {\cal T} \partial_{\lambda} L\right) d\lambda.
\end{align}
 This leads to a modified Smarr formula
 \begin{align}
 (n+1)M = (n+2)T_H S + {\cal T} L+2\alpha \left(\partial_{\alpha} M
-T_H \partial_{\alpha} S- {\cal T} \partial_{\alpha} L \right).\label{eq:smarr}
 \end{align}

  \section{Phase and stability of static black strings} \label{sec:staticphase}
  
  Now, solving the large $D$ effective theory up to NLO, we study the phase and stability of uniform/non-uniform black strings. In particular, we focus on the weakly deformed phase which can be solved analytically by the perturbative method.
  
 \subsection{Stability of uniform black string}
We start with examining the stability of the uniform phase refining the leading order result in ref.~\cite{Chen:2017rxa}. 
With the linear perturbation from the uniform solution
\begin{align}
 m(t,z) = 1 +  \veps\, m_1 e^{\Omega t} \cos (kz),\quad p(t,z) =\veps \, p_1 e^{\Omega t} \sin(kz),
\end{align}
eq.~(\ref{eq:NLOeq-conserve}) at $\ord{\veps}$ reduces to
\begin{align}
   A_{11} \, m_1+A_{12} \, p_1 =0,\quad A_{21} \, m_ 1+ A_{22} \, p_1 =0,
\end{align}
where $A_{ij}$ are functions of $\Omega$ and $k$. The degeneracy condition ${\rm det}(A_{ij}) =0$
leads to the dispersion relation as
\begin{align}
 \Omega = \Omega_{\rm LO} + \fr{n}\Omega_{\rm NLO},
\end{align}
where
\begin{align}
 \Omega_{\rm LO} = -\frac{1+2\alpha+2\alpha^2}{(2\alpha+1)(\alpha+1)}k^2 \pm  \frac{k \sqrt{4 \alpha ^4+8 \alpha ^3+4 \alpha +\alpha ^2 \left(k^2+7\right)+1}}{(2\alpha+1)(\alpha+1)},
\end{align}
\begin{align}
\Omega_{\rm NLO}& = -\frac{2 \alpha  \left(2 \alpha ^2+2 \alpha +1\right) k^4 \arctan (2 \alpha +1)}{(\alpha +1)^3 (2
   \alpha +1)^2}  -\frac{\alpha  \left(2 \alpha ^2+2 \alpha +1\right) k^4 \log(2\alpha^2+2\alpha+1)}{(\alpha +1)^3 (2
   \alpha +1)^3}\nonum
   &-\frac{2 \alpha  \left(2 \alpha ^2+2 \alpha +1\right) k^4 \log(1+\alpha)}{(\alpha +1)^3
   (2 \alpha +1)}-\frac{\left(40 \alpha ^6+104 \alpha ^5+112 \alpha ^4+60 \alpha ^3+20 \alpha ^2+6 \alpha
   +1\right) k^2}{(\alpha +1)^2 (2 \alpha +1)^2 \left(2 \alpha ^2+2 \alpha +1\right)}
\nonum
& +\frac{\alpha  (8 \alpha +\pi +4) \left(2 \alpha ^2+2 \alpha +1\right) k^4}{2 (\alpha +1)^3 (2 \alpha
   +1)^2}\pm
   \frac{k \sqrt{4 \alpha ^4+8 \alpha ^3+7 \alpha ^2+4 \alpha +\alpha ^2 k^2+1}}{4 (\alpha +1)^3 (2
   \alpha +1)^3}\nonum
&   \times \left[\frac{4 \left(4 \alpha ^3+6 \alpha ^2+4 \alpha +1\right) k^3 \arctan(2 \alpha +1) \left(4 \alpha
   ^4+16 \alpha ^3+8 \alpha +\alpha ^2 \left(19-2 k^2\right)+1\right)}{4 \alpha ^4+8 \alpha ^3+4
   \alpha +\alpha ^2 \left(k^2+7\right)+1}\right.\nonum
   &\left.-\frac{4 (2 \alpha +1)^2 k^3\log(\alpha+1)
   \left(-8 \alpha ^5+2 \alpha +4 \alpha ^4 \left(k^2-5\right)+2 \alpha ^3 \left(2
   k^2-9\right)+\alpha ^2 \left(2 k^2-5\right)+1\right)}{4 \alpha ^4+8 \alpha ^3+4 \alpha +\alpha ^2
   \left(k^2+7\right)+1}\right.\nonum
   & \left.+\frac{2\left(2 \alpha ^2+2 \alpha +1\right) k^3 \log(2\alpha^2+2\alpha+1) \left(4 \alpha
   ^4+16 \alpha ^3+8 \alpha +\alpha ^2 \left(19-2 k^2\right)+1\right)}{4 \alpha ^4+8 \alpha ^3+4
   \alpha +\alpha ^2 \left(k^2+7\right)+1}\right.\nonum
   &\left. +\frac{(2 \alpha +1) k}{4 \alpha ^4+8 \alpha ^3+4 \alpha +\alpha ^2 \left(k^2+7\right)+1}\left(2 (\alpha +1)^2 \left(8 \alpha ^5-28 \alpha ^4-46 \alpha ^3-19 \alpha ^2-4 \alpha -1\right)\right.\right.\nonum
   &\left.\left.+2 \alpha   ^2 (8 \alpha +\pi +4) \left(2 \alpha ^2+2 \alpha +1\right) k^4-(\alpha +1) \left(\pi  \left(8
   \alpha ^5+32 \alpha ^4+42 \alpha ^3+28 \alpha ^2+9 \alpha +1\right)\right.\right.\right.\nonum
&  \left. \left.\left.-4 \left(16 \alpha ^6+24
   \alpha ^5+12 \alpha ^4-9 \alpha ^3-8 \alpha ^2+\alpha +1\right)\right) k^2
   \right)\right].
\end{align}
The threshold wave number $\Omega(k_{\rm GL})=0$ for the instability up to NLO is given by 
\begin{align}
&k_{\rm GL} =1-\frac{k_1}{n},\label{eq:kgl-1}
\end{align}
where $k_1$ is written as $k_1 = c_{01}/2$ in terms of $c_{01}$ in~(\ref{eq:static-eq-cfs01}).

\begin{figure}[t]
\begin{center}
\includegraphics[width=7cm]{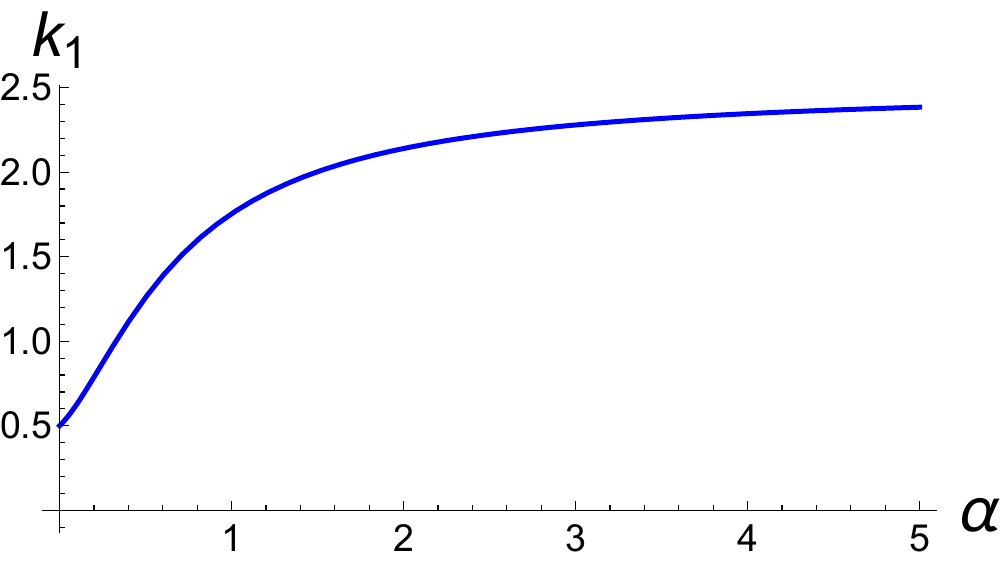}
\caption{$\alpha$-dependence of $k_1$ \label{fig:dkGL}}
\end{center}
\end{figure}

\medskip
As shown in ref.~\cite{Chen:2017rxa}, the threshold scale for the instability is the same as in GR at the leading order. 
With the NLO correction, one can see the dependence on the coupling constant $\alpha$ in the threshold. 
As $\alpha$ grows, $k_{\rm GL}$ monotonically decreases, and then it has a finite limit at $\alpha \to \infty$ (figure \ref{fig:dkGL}),
\begin{align}
 k_{\rm GL}\bigr|_{\alpha = 0} =1-\fr{2n} \quad \longrightarrow \quad k_{\rm GL}\bigr|_{\alpha\to \infty} = 1-\frac{5}{2n}.
 \end{align}
%
%\if0
%\fi
%
The corresponding periodicity is given by
 \begin{align}
 L_{\rm GL} = \frac{2\pi}{k_{\rm GL}}= 2\pi r_0 \left(1 + \frac{k_1}{n}\right)+\ord{n^{-2}},\label{eq:def-LGL}
 \end{align}
 where we reinstated the radius scale.
 Therefore, requiring
 longer wavelength for the instability, 
uniform black strings are stabilized by the GB correction.  
Note that this is a consistent result with the numerical analysis for $D=9$ pure GB black strings, which shows the smaller threshold wavenumber than for GR~\cite{Giacomini:2015dwa,Giacomini:2016ftc}. 
Interestingly, this is contrary to the result in $5\leq D\leq 8$ where the instability needs larger $k_{\rm GL}$ for small $\alpha$~\cite{Brihaye:2010me,Henriquez-Baez:2022bfi}, which implies a critical dimension on the $\alpha$-dependence of $k_{\rm GL}$. To confirm such a critical dimension, one needs to know the NNLO correction to eq.~(\ref{eq:kgl-1}).

\subsection{Weakly non-uniform black string} 
The static non-uniform black string can be constructed by the perturbative expansion from the static uniform black string in the small non-uniformity by starting from the onset of the instability. 
As in GR~\cite{Suzuki:2015axa,Emparan:2018bmi}, we assume the following expansion in the static equation~(\ref{eq:static-eq}) 
\begin{align}
   \cR(z) = \sum_{i=1}\veps^i \mu_{i} \cos(i k z),\quad \mu_i = \sum_{j=0}\sum_{k=0} \mu_{i,j,k}\veps^j n^{-k}, \label{eq:nubs-pertsol}
\end{align}
where without loss of generality, we can set $\mu_1=1$ by the redefinition of small parameter $\veps$.
The first three terms are given by
\begin{align}
&\mu_0 =-\veps ^2 \left(\frac{1}{4}-\frac{c_{02}+c_{20}}{2 n}\right)+\veps ^4 \left(\frac{1}{72}-\frac{19 c_{02}+24 c_{20}+9 c_{21}}{288 n}\right)+\ord{\veps^6},\nonum
&\mu_2=   -\frac{1}{12}+\frac{c_{20}-c_{02}}{6   n}+\veps ^2 \left(\frac{5}{1728}+\frac{13 c_{21}-5 c_{02}-10   c_{20}}{576 n}\right)+\ord{\veps^4},\nonum%\veps ^4 \left(\frac{280 c_{02}+245   c_{20}-731 c_{21}}{414720 n}-\frac{49}{829440}\right),\nonum
&\mu_3 =\frac{1}{96}+\frac{3 c_{02}-4 c_{20}+3 c_{21}}{96 n}-\veps ^2 \left(\frac{29}{46080}-\frac{17 c_{02}+116 c_{20}-195 c_{21}}{23040   n}\right)+\ord{\veps^4},%+ \veps ^4 \left(\frac{-11117 c_{02}-15684 c_{20}+57651 c_{21}}{66355200   n}+\frac{1307}{66355200}\right),
\end{align}
where $c_{ij}$ are shown in eq.~(\ref{eq:static-eq-cfs}).
The wavenumber is also modified from $k_{\rm GL}$ 
\begin{align}
&\frac{k}{k_{\rm GL}}=1
-\frac{\veps^2}{24} \left(1-\frac{5 c_{02}+4 c_{20}-3 c_{21}}{n}\right)
 +   \veps^4 \left(\frac{19}{6912}+\frac{-95 c_{02}-76 c_{20}+33 c_{21}}{3456 n}\right)+\ord{\veps^6}.
\end{align}
The small amplitude $\veps$ is related to
the non-uniformity parameter~(\ref{eq:def-lambda}) by eq.~(\ref{eq:nubs-pertsol})
\begin{align}
&\lambda =\fr{2}\left(\frac{e^{\cR(0)}}{e^{\cR(\pi)}}-1\right)\nonum
&=\veps +\veps^2+\frac{\veps ^3}{96}
   \left(65+\frac{3 c_{02}-4 c_{20}+3 c_{21}}{n}\right)+ \frac{\veps ^4}{48} \left(17+\frac{3 c_{02}-4 c_{20}+3 c_{21}}{n}\right)+\ord{\veps^5}.\label{eq:def-lam}
\end{align}
The periodicity is, then, expanded in terms of $\lambda$,
\begin{align}
&\frac{L}{L_{\rm GL}} =1+\frac{\lambda ^2}{24} \left(1-\frac{\ell_1}{n}\right)
-\frac{\lambda ^3}{12} \left(1-\frac{\ell_1}{ n}\right)+\lambda ^4 \left(\frac{1043}{6912}-\frac{515 \ell_1+72}{3456 n}\right)+\ord{\lambda^5},
\end{align}
where $L_{\rm GL}$ is given by eq.~(\ref{eq:def-LGL}) and $\ell_1$ is written as
\begin{align}
&\ell_1 :=5 c_{02}+4 c_{20}-3 c_{21}
=\frac{48 \alpha ^3+76
   \alpha ^2+(40+\pi ) \alpha +16}{2(\alpha+1)(2\alpha^2+\alpha+1)}\nonum
&  \qquad -\frac{\alpha  \left(\log \left(2 \alpha ^2+2 \alpha +1\right)+2 (2 \alpha +1)^2 \log (\alpha +1)+2(2 \alpha   +1) \arctan(2 \alpha +1)\right)}{(\alpha +1) (2 \alpha +1) \left(2 \alpha ^2+\alpha +1\right)}.\label{eq:def-ell1}
\end{align}

 \subsection{Thermodynamics}
 Using the perturbative solution of the non-uniform black string constructed in the previous section, we can obtain the thermodynamic variables as the expansion in $\lambda$,
\begin{align}
&\frac{M}{M_{\rm GL}} =1+\frac{\lambda ^2}{24}
   \left(1-\frac{\ell_1-12}{n}\right)-\frac{\lambda ^3}{12}
   \left(1-\frac{\ell_1-12}{n}\right)
+\lambda ^4 \left(\frac{5448-443 \ell_1}{3456 n}+\frac{971}{6912}\right)+\ord{\lambda^5},\\
& \frac{S}{S_{\rm GL}}=1+\frac{\lambda ^2}{24}
   \left(1-\frac{\ell_1-12}{n}\right)-\frac{\lambda ^3}{12}
   \left(1-\frac{\ell_1-12}{n}\right)
+\lambda ^4 \left(\frac{5448-443 \ell_1}{3456 n}+\frac{971}{6912}\right)+\ord{\lambda^5},\\
&\frac{{\cal T}}{{\cal T}_{\rm GL}}=
1-\left(1+\frac{9-\ell_1}{n}\right) \frac{\lambda^2}{2}+\left(1+\frac{9-\ell_1}{n}\right) \lambda
   ^3-\left(\frac{121}{72}+\frac{55 (9-\ell_1)}{36 n}\right) \lambda ^4+\ord{\lambda^5},
\end{align}
where $\ell_1$ is shown in eq.~$(\ref{eq:def-ell1})$ and the critical values for uniform strings, where non-uniform strings branch off, are given by
\begin{align}
&M_{\rm GL} =(1+\alpha) r_0^{n+1} \left(1+\frac{(\alpha+1)  \ell_1-7 \alpha-5 }{2 n(1+\alpha)}\right),\\
&S_{\rm GL} = (1+2\alpha)r_0^{n+2}  \left(1+\frac{(2 \alpha+1) 
  \ell_1 -10 \alpha-7}{2 n(1+2\alpha)}\right),\\
&{\cal T}_{\rm GL} = \frac{r_0^n }{n}\frac{1+\alpha +2 \alpha ^2}{1+2 \alpha }\left[1+\fr{n}\left(\frac{\alpha  \left(4 \alpha ^2-8 \alpha -3\right)}{(1+2 \alpha ) \left(1+\alpha +2 \alpha  ^2\right)}\right.\right.\nonum
&\left.\left.\hspace{3cm}+\frac{\arctan(1+2 \alpha )-\pi/4-\log (1+\alpha )}{1+ \alpha }+\frac{\log \left(1+2
   \alpha +2 \alpha ^2\right)}{2(2\alpha+1)(\alpha+1)}\right)\right],
\end{align}
and
\begin{align}
& T_H = T_{H,{\rm GL}} =    \fr{r_0} \left(\frac{1+\alpha}{1+2\alpha} - \frac{\alpha(1+4\alpha)}{n(1+2\alpha)^2}\right).
\end{align}
In the above, we  re-introduce the radius scale.
Note that the same $\lambda$-dependence that appears in the mass and entropy provides
 just an approximate relation up to NLO.\footnote{With the form of $L_{\rm GL}, M_{\rm GL}, S_{\rm GL}, {\cal T}_{\rm GL}$ and the assumption $\partial_\alpha (M/M_{\rm GL})=\ord{n^{-1}}=\partial_\alpha (S/S_{\rm GL})$, the Smarr formula~(\ref{eq:smarr}) shows $M/M_{\rm GL}-S/S_{\rm GL}=\ord{n^{-2}}$. This is because the $\alpha$-dependence does not appear in the static equation at the leading order.}

For these perturbation formula, we can confirm that the Smarr relation~(\ref{eq:smarr}) and 
\begin{align}
 \partial_\lambda M =  T_H \partial_\lambda S+ {\cal T} \partial_\lambda L
\end{align}
hold up to $\ord{\lambda^{10}}$ and NLO in $1/n$.

In the literature~\cite{Sorkin:2006wp}, the relative binding energy is also defined as a useful scale invariant quantity
\begin{align}
& \tau := \frac{L{\cal T}}{{\cal M}} \nonum
&\quad = \tau_{\rm GL} \left[1-\left(1-\frac{\ell_1-10}{n}\right) \frac{\lambda ^2}{2}+\left(1-\frac{\ell_1-10}{n}\right)
   \lambda ^3-\left(\frac{481}{288}+\frac{2171-217 \ell_1}{144 n}\right) \lambda ^4+\ord{\lambda^5}\right],
\end{align}
where
\begin{align}
 \tau_{\rm GL} = \frac{L_{\rm GL} {\cal T}_{\rm GL}}{M_{\rm GL}} = \fr{n}\frac{2\alpha^2+\alpha+1}{(2\alpha+1)(\alpha+1)}- \frac{1+6\alpha+15\alpha^2+8\alpha^3-4\alpha^4}{n^2(1+\alpha)^2(1+2\alpha)^2}.
\end{align}
For the uniform solution, one can check that the large $\alpha$ limit of the relative binding energy and temperature matches with the result by the large $\alpha$ approximation up to NLO in $1/n$~\cite{Suzuki:2022eaz}
\begin{align}
 \tau_{\rm GL} \to \fr{n}+\fr{n^2} \simeq \fr{n-1},\quad \frac{n}{4\pi} T_{H,{\rm GL}} \to \frac{n-2}{8\pi r_0} \quad ( \alpha \to \infty). 
\end{align}

\subsection{Stability of weakly non-uniform black string}
By considering the dynamical perturbation to the static solution~(\ref{eq:nubs-pertsol}) in eq.~(\ref{eq:NLOeq-conserve}),
we study the change of the fundamental modes
\begin{align}
 m(t,z) = e^{\cR(z)} \left( 1+ \delta \, e^{\Omega t} f(z) \right),\quad
  p(t,z) = p_{s}(z) \left(1+ \delta \, e^{\Omega t}g(z)\right),
\end{align}
where $\delta$ is another small parameter and $p_{s}(z)$ corresponds to that of the static solution~(\ref{eq:static-p}). We assume the mode functions are decomposed by the Fourier modes in the periodicity of the static solution,
\begin{align}
f(z) = \sum_{i=0} \veps^i \nu_i \cos (i k z),\quad g(z) = \sum_{i=0} \veps^i \bar{\nu}_i \sin (i k z).
\end{align}
We focus on the fundamental mode which goes back to the zero mode $\Omega \to 0$ for $\veps \to 0$\footnote{All other modes are stable at $\veps\to 0$ and then, will not give the instability in the perturbative regime.}. 
Here, we do not give the detail of the derivation, but find that the growth rate becomes
\begin{align}
\Omega= - \frac{2\alpha^2+\alpha+1}{2\alpha^2+2\alpha+1}\left(1- \frac{ \Omega_1}{n}\right)\frac{\lambda^2}{12}+\ord{\lambda^3},\label{eq:nubs-omega}
\end{align}
where $\veps$ is replaced with $\lambda$ by using eq.~(\ref{eq:def-lam}) and
\begin{align}
&  \Omega_1 =  \fr{(\alpha+1)(2\alpha^2+2\alpha+1)}\left(
 32 \alpha ^3+57 \alpha ^2+(26+\pi ) \alpha+9+\frac{2 \alpha ^3-2 \alpha
   ^2-\alpha }{\left(2 \alpha ^2+2 \alpha
   +1\right)^2} +\frac{1}{2 \alpha
   +1}\right.\nonum
&\left. \qquad   -4 \alpha  \arctan (2 \alpha +1)
   -4 (2 \alpha   +1) \alpha  \log(\alpha+1)
   -\frac{2 \alpha  \log(2\alpha^2+2\alpha+1)}{2   \alpha +1}\right).
\end{align}
Therefore, for large enough $n$, we obtain $\Omega<0$, i.e. the non-uniform phase is linearly stable.
For smaller $n$, the stability of the non-uniform phase depends on the value and signature of $\Omega_1$.

\subsection{Critical dimensions}
In the GR case, 
it is known that the black string admits a critical dimension $n_*$, such that for $n>n_*$ the non-uniform phase is stable, and for $n<n_*$ unstable~\cite{Sorkin:2004qq}, where the critical dimension appears in the mass for fixed $L$ as well as in the entropy difference between the non-uniform and uniform phase of the same mass. We can see the critical dimension from the mass for fixed $L$ expanding up to $\ord{\lambda^2}$
\begin{align}
 \frac{M}{L^{n+1}} \simeq \frac{M_{\rm GL}}{L_{\rm GL}^{n+1}}\left[1 
 + \frac{n \lambda^2}{24} \left(1-\frac{\ell_1}{n}\right)\right],
\end{align}
where $\ell_1$ is given by eq.~(\ref{eq:def-ell1}) and we assume the deformation parameter is sufficiently small $\lambda^2 \ll 1/n$.
This estimates the critical dimension in the mass as
\begin{align}
 n_{*,{\rm M}}=\ell_1\quad (D_{*,{\rm M}}=4+\ell_1).
\end{align}
To determine the critical dimension in the entropy difference at large $D$, we need to solve the $\ord{n^{-2}}$ correction. Hence, we do not discuss this.

Similarly, the critical dimension also appears in the dynamical stability~(\ref{eq:nubs-omega}), where we obtain
\begin{align}
 n_{*,{\rm D}}=\Omega_1,\quad (D_{*,{\rm D}}=4+\Omega_1).
\end{align}

\begin{figure}[t]
\begin{center}
\includegraphics[width=10cm]{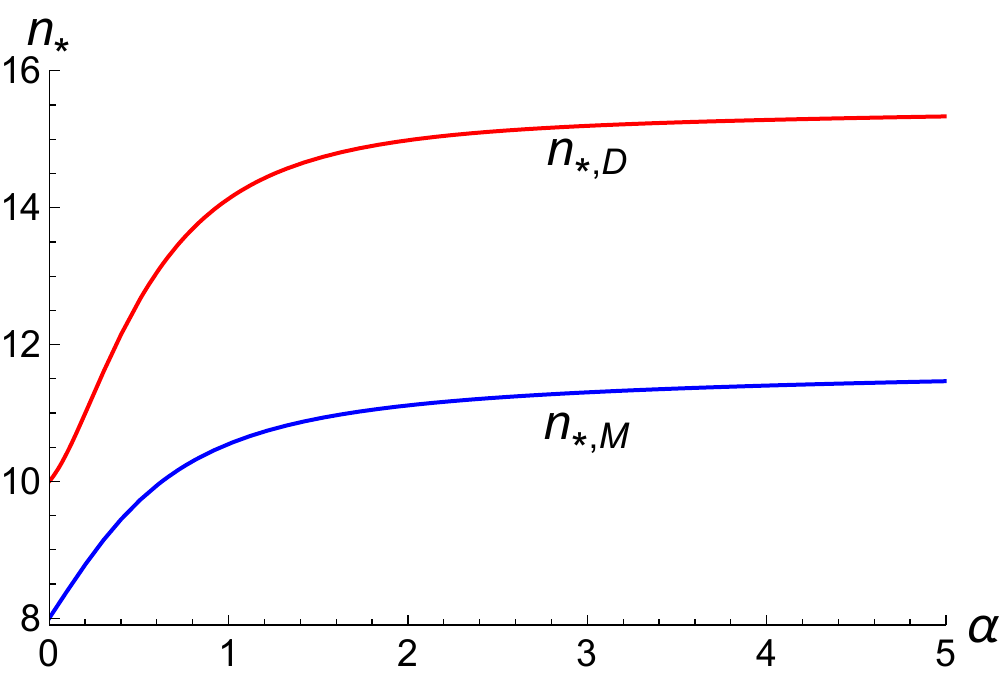}
\caption{$\alpha$-dependence of the critical dimensions in the mass and stability\label{fig:Ncrits}}
\end{center}
\end{figure}

As shown in figure~\ref{fig:Ncrits}, the critical dimensions for both the mass and dynamical stability increase as the coupling constant $\alpha$ grows.
This indicates that we need a slightly higher dimension for the stable non-uniform phase with a large coupling constant $\alpha$.

As discussed in ref.~\cite{Emparan:2018bmi}, in the GR case, the NLO result determines $n_*$ only up to $\ord{n^{0}}$, that is 
\begin{align}
n = n_* \left(1+\ord{1/n_*}\right) = n_* + \ord{n_*^0}.
\end{align}
\medskip
In the GR case, these critical dimensions approaches to the similar values that give the same threshold for the integer dimension with higher order corrections in $1/n$.
%\if0
%\fi
With the GB correction, it is unclear whether or not $n_{*,{\rm M}}$ and $n_{*,{\rm D}}$ give the same criticality. 
To confirm $n_{*,{\rm M}}=n_{*,{\rm D}}$ for every $\alpha$, one might need a full numerical study with non-integer $n$. If they admit a slight difference, one can take a value of $\alpha$ to satisfy either of (i) $n_{*,{\rm M}}<n<n_{*,{\rm D}}$ or (ii) $n_{*,{\rm D}}<n<n_{*,{\rm M}}$. From figure~\ref{fig:Ncrits}, it seems to be the case (i), in which the non-uniform phase will be dynamically unstable, but thermodynamically stable. However, we cannot expect any such physical system. Fortunately, the GR result up to $\ord{n^{-3}}$~\cite{Emparan:2018bmi} estimates $n_{*,{\rm M,GR}} = 9.93$ and $n_{*,{\rm D,GR}}=9.62$ supporting the case (ii),
in which the non-uniform phase will be dynamically stable but thermodynamically unstable, i.e. metastable.

Another possibility is that both $n_{\rm *,D}(\alpha)$ and $n_{\rm *,M}(\alpha)$ simply converge within $9<n<10$ with higher order corrections in $1/n$ uniformly for any $\alpha$, which has the same threshold for integer dimensions as in GR.

\section{Conclusion}\label{sec:sum}
In this article, we have developed the analytical studies on the non-linear dynamics of black strings in Einstein-Gauss-Bonnet (EGB) theory, using the large $D$ effective theory approach. 
Extending the earlier work~\cite{Chen:2017rxa}, we have obtained the dynamical effective theory up to the next-to-leading order (NLO) in the $1/D$-expansion. Remarkably, with the NLO correction, we have shown that the Iyer-Wald entropy of the dynamical deformation 
is subject to the second law within the large $D$ effective theory. Using the entropy and mass formulae up to NLO, we have also determined the entropy functional for the leading order theory.

\medskip
Solving the effective equation, we have studied the phase and stability of uniform and non-uniform black strings up to NLO.
We have found that the Gregory-Laflamme instability of uniform solutions requires longer wavelength than in General Relativity (GR).
The threshold wavelength grows as a function of the Gauss-Bonnet (GB) coupling constant $\alpha_{\rm GB}$, and finally ends at a finite limit for $\alpha_{\rm GB}\to\infty$. This indicates that the GB correction stabilize uniform black strings. 

\medskip
From the onset of the instability, we have also constructed static non-uniform black strings by perturbing static uniform black strings in the weakly-deformed regime and have studied the thermodynamics and dynamical stability.  
We have estimated the critical dimensions for both the mass with periodicity fixed and the dynamical stability within the accuracy of $\ord{n^0}$, and have indicated that both of them are increasing functions of $\alpha_{\rm GB}$. 
As the threshold of the instability, the critical dimensions approach to certain finite values at the limit $\alpha_{\rm GB} \to \infty$. This shows non-uniform black strings are rather destabilized by the GB correction, contrary to uniform black strings.
With the GB correction, a slightly higher dimension is required to admit the stable non-uniform phase.
We have also agued that the difference between the thermodynamical and dynamical critical dimensions leads to the metastable non-uniform phase in a certain dimension with a certain value of $\alpha_{\rm GB}$.

\medskip
A simple extension of this work is finding higher order corrections in the $1/D$-expansion, which nevertheless will require much harder works to find the metric solution.
With higher order corrections in the $1/D$-expansion, one can determine the critical dimensions more accurately.
This will predict the existence of the metastable non-uniform phase, more correctly.
It will also be interesting to explore such metastable phase directly in the fully numerical analysis.

\medskip
The large $D$ effective theory approach will be also applicable to the Einstein-Lovelock theories or more general higher curvature theories,
in which the second law is only shown for the perturbation around the stationary black holes~\cite{Wall:2015raa,Hollands:2022fkn}.
It will be possible to verify the second law beyond the perturbative level in those theories almost in parallel by using the large $D$ effective theory approach, but one must replace the Iyer-Wald entropy with the Iyer-Wald-Wall entropy~\cite{Wall:2015raa}.

\section*{Acknowledgement}
This work is supported by Toyota Technological Institute Fund for Research Promotion A.
RS was supported by JSPS KAKENHI Grant Number~JP18K13541.
ST was supported by JSPS KAKENHI Grant Number 21K03560. 

\appendix

\section{$1/D$-expansion of EGB equation}
In this section, we explain the derivation of the metric solution in the $1/D$-expansion.
Instead of eq~(\ref{eq:EGB-eom}), we expand an equivalent equation
\begin{align}
{\cal E}_{\mu\nu} := R_{\mu\nu}+ \alpha_{\rm GB} \tilde{H}_{\mu\nu}=0,\label{eq:EGB-eom-2}
\end{align}
where
\begin{align}
&\tilde{H}_{\mu\nu} = -\fr{D-2}\cL_{\rm GB}g_{\mu\nu}+2RR_{\mu\nu}-4R_{\mu\alpha}R^\alpha{}_\nu - 4 R_{\mu\alpha\nu\beta}R^{\alpha\beta} + 2 R_{\mu\alpha\beta\gamma}R_{\nu}{}^{\alpha\beta\gamma}.
\end{align}
We denote the sphere component as
\begin{align}
 \cE_{ij} = \cE_{\Omega} \gamma_{ij},
\end{align}
where $\gamma_{ij}$ is the metric for $S^{n+1}$. With the assumption $\alpha := n^{2}\alpha_{\rm GB}=\ord{1}$ and the new radial coordinate ${\sf R}:=r^n$, the sphere component gives the equation for $A$ at the leading order
\begin{align}
 \cE_\Omega/n = \sR(2\alpha A_0-2\alpha-1)\partial_{\sR} A_0 + (A_0-1)(\alpha A_0-\alpha-1) + \ord{1/n}.
 \label{eq:LOeq-A0}
\end{align}
The solution is easily found as eq.~(\ref{eq:LOsolA}). Given the solution of $A_0$, the $1/n$ expansion of $\cE_{rz}, \cE_{zz},\cE_{rr}$ 
lead to the radial ODE for $C,G$ and $U$ at the leading order, respectively,
\begin{align}
&2 \cE_{rz}/n =  \partial_{\sR} \left( \sR^2(2\alpha A_0-2\alpha-1)\partial_{\sR} C_0\right) + \ord{1/n},\label{eq:LOeq-C0}\\
&2\cE_{zz} = \partial_\sR \left[\sR^2 A_0(2\alpha\sR \partial_{\sR} A_0+2\alpha A_0-2\alpha-1)\partial_\sR G_0)\right]\nonum
&\quad - \partial_\sR \left[ 
2\alpha  \sR   (A_0 -1) \left(2 \sR \partial_\sR A_0 +A_0 -1\right)
-2\sR \partial_z C_0    \left(2 \alpha  \sR \partial_\sR A_0 +2 \alpha  A_0 -2 \alpha -1\right)\right.\nonum
&\quad \quad\left.  +4 \alpha  \sR^2 \partial_z A_0  \partial_\sR C_0 - 2\alpha  \sR^3 A_0   (\partial_\sR C_0) ^2\right]+ \sR^2 (\partial_\sR C_0) ^2 (2 \alpha  A_0 -2 \alpha -1)+\ord{1/n},\label{eq:LOeq-G0}\\
&\cE_{rr}/n = -{\sf R} ( 2\alpha A_0-2\alpha-1) \partial_{\sR} \left(U_0 - \frac{\sR}{2} \partial_{\sR} G_0\right) - \alpha \sR^2(\partial_{\sR} C_0)^2+\ord{1/n},\label{eq:LOeq-U0}
\end{align}
where the latter two also need the solution for $C_0$ as sources.

Substituting the $1/n$-expansion of the metric~(\ref{eq:metricexp}) to $\cE_\Omega,\ \cE_{rz},\ \cE_{zz}$ and $\cE_{rr}$, one can obtain $i$-th order evolution equation for $i\geq 1$, which is rewritten in terms of the auxiliary variable $X$~(\ref{eq:def-X}) as,
\begin{subequations}\label{eq:nloeq}
\begin{align}
&\partial_X \left( \frac{X}{1-X^2} A_{i} \right) = {\cal S}_A^{(i)},\label{eq:nloeq-A}\\
&\partial_X^2 C_i = {\cal S}_C^{(i)}\label{eq:nloeq-C},\\
&\partial_X [(1+X^{-2})(1+2\alpha-X) \partial_X G_i] = {\cal S}_G^{(i)}\label{eq:nloeq-G},\\
&\partial_X \left( U_i -\frac{1-X^2}{4X} \partial_{X} G_i \right) = {\cal S}_U^{(i)}\label{eq:nloeq-U},
\end{align}
\end{subequations}
where the sources ${\cal S}_A^{(i)}$ and ${\cal S}_C^{(i)}$ only contain lower order solutions, while  ${\cal S}_U^{(i)}$ and ${\cal S}_G^{(i)}$
further include $A_i$ and $C_i$ as in eqs.~(\ref{eq:LOeq-G0}) and (\ref{eq:LOeq-U0}).

For other components, one can find that $\cE_{tt}$ and $\cE_{tr}$ gives the condition degenerate to $\cE_\Omega$ at leading order
\begin{align}
 2 g_{tt}^{-1} \cE_{tt}/n^2 \simeq 2\cE_{tr}/n^2 \simeq \sR \partial_\sR \cE_\Omega/n =  r \partial_r \cE_\Omega/n^2.
 %2n^{-2} \cE_{tr} \simeq 2 \alpha  \sR^2 (\partial_\sR A)^2
% +(2 \alpha  A-2  \alpha -1)\left(2\sR \partial_\sR A+R^2 \partial_\sR^2 A \right),
\end{align}
%which is identical to $\sR \partial_\sR \cE_\Omega = n^{-1} r \partial_r \cE_\Omega$ at the leading order.
Eliminating the leading order terms in $\cE_{tt}$ and $\cE_{tr}$ with $\cE_\Omega$, we obtain the conditions correspond to the Hamiltonian constraint
\begin{align}
 {\cal H} := 2 n^{-1} \cE_{tr}- n^{-1} r \partial_r \cE_\Omega =0 ,\label{eq:constH}
\end{align}
and temporal part of the vector constraint,
\begin{align}
 {\cal V}_t := 2n^{-1}  ( \cE_{tt}-g_{tt} \cE_{tr}) =0.\label{eq:constT}
\end{align}
Similarly, the leading order part of $\cE_{tz}$ is degenerate to the combination of $\cE_{rz}$ and $\cE_\Omega$, and then, the subtraction of it leads to the condition corresponds to the constraint along the string direction
\begin{align}
 {\cal V}_z := 2\cE_{tz}- 2g_{tt} \cE_{rz}-\partial_z \cE_\Omega=0 .\label{eq:constZ}
\end{align}
One can find that ${\cal V}_t=0$ and ${\cal V}_z=0$  produce the effective equations if the metric solutions are substituted, while $\cH=0$ is trivially satisfied without extra condition.

\section{Necessary functions for higher order integrals}
\label{sec:funcs}
Here we introduce several functions which appear in the higher order solution of metric functions.
These functions are necessary to perform integrals involving $\arctan(x)$ and $\log(1+x^2)$.

\paragraph{Clausen's function}
Clausen's functions are related to the real and imaginary parts of $\Li_k(e^{i\theta})$
\begin{align}
 \Li_{2m}(e^{i\theta})={\rm Sl}_{2m}(\theta)+i{\rm Cl}_{2m}(\theta),\quad
  \Li_{2m+1}(e^{i\theta})={\rm Cl}_{2m}(\theta)+i{\rm Sl}_{2m}(\theta)
\end{align}
where ${\rm Cl}_k(\theta)$ and ${\rm Sl}_k(\theta)$ are also defined as the Fourier series
\begin{align}
& {\rm Cl}_{2m}(\theta) = \sum_{k=1}^\infty \frac{\sin(k \theta)}{k^{2m}},\quad
  {\rm Sl}_{2m}(\theta) = \sum_{k=1}^\infty \frac{\cos(k \theta)}{k^{2m}},\\
&    {\rm Cl}_{2m+1}(\theta) = \sum_{k=1}^\infty \frac{\cos(k \theta)}{k^{2m+1}},\quad
        {\rm Sl}_{2m+1}(\theta) = \sum_{k=1}^\infty \frac{\sin(k \theta)}{k^{2m+1}}.
\end{align}
By definition, these functions have the periodicity of $2\pi$.
The ${\rm Sl}$-type functions are expressed by the Bernoulli polynomials for $0 \leq \theta \leq 2\pi$
\begin{align}
  {\rm Sl}_{2m} (\theta) = \frac{(-1)^{m-1}(2\pi)^{2m}}{2(2m)!}B_{2m}\left(\frac{\theta}{2\pi}\right),\quad
    {\rm Sl}_{2m-1} (\theta) = \frac{(-1)^{m}(2\pi)^{2m-1}}{2(2m-1)!}B_{2m-1}\left(\frac{\theta}{2\pi}\right).
\end{align}
${\rm Cl}$-type functions also satisfies the double angle formula
\begin{align}
 {\rm Cl}_{m+1} ( 2\theta) = 2^m [{\rm Cl}_{m+1} (\theta)+
 (-1)^m {\rm Cl}_{m+1}(\pi-\theta)].
\end{align}
${\rm Cl}_2$ is used to express following integrations
\begin{align}
&\int \frac{\log x}{1+x^2} dx = - \fr{2} {\rm Cl}_2(2 \arctan x)
+ \fr{2} {\rm Cl}_2(2 \arctan x+\pi),\\
&\int \frac{\log (x\pm 1)}{1+x^2} dx = - \fr{2} {\rm Cl}_2(2 \arctan x \pm \pi/2) + \fr{2} {\rm Cl}_2(2 \arctan x+\pi)
+ \fr{2}\arctan x \log 2,\\
& \int \frac{\log(1+x^2)}{1+x^2} dx = {\rm Cl}_2 (2\arctan x+\pi)+ 2  \arctan x \log 2.
\end{align}

\paragraph{Complex polylogarisms}
Clausen's functions are the polylogarism on the unit circle in the complex plain.
We also need the polylogarism on other trajectories in the complex plain.
We define the following real functions as the real and imaginary parts of corresponding polylogarism
\begin{align}
&  \ReLiA(x) = {\rm Re}\, \Li_2\left(\frac{1-i}{2}(1+x)\right),\quad
   \ImLiA(x) = {\rm Im}\, \Li_2\left(\frac{1-i}{2}(1+x)\right)\\
  &   \ReLiB(x) = {\rm Re}\, \Li_2\left(\frac{x-1-2\alpha}{i+x}\right),\quad
       \ImLiB(x) = {\rm Im}\, \Li_2\left(\frac{x-1-2\alpha}{i+x}\right).
\end{align}
It turns out that we do not need $\ImLiA(x)$ since it is expressed by ${\rm Cl}_2$'s
\begin{align}
&\ImLiA(x) =  \fr{2}{\rm Cl}_2(2\arctan x+\pi)-\fr{2}{\rm Cl}_2\left(2 \arctan x +\frac{\pi}{2}\right) \nonum
&\qquad-\left(\arctan x+\frac{\pi}{4}\right)\log (1+x)+ \fr{2}\left(\arctan x +\frac{\pi}{4}\right)\log 2 - \frac{K}{2},
\end{align}
where $K\approx 0.916$ is Catalan's constant.

These functions are used to express following integrations 
\begin{align}
&\int \frac{x \log(x\pm 1)}{1+x^2}dx = \ReLiA(\pm x)-\fr{2} \log(x+1) \log \left(\frac{x^2+1}{2}\right),\\
& \int \frac{\log(1+x^2)}{x-1-2\alpha} dx = 2 \ReLiB(x)-\arctan^2 x + \fr{4}\log(1+x^2)^2 + 2 \arctan x \arctan(1+2\alpha) \nonum
&\hspace{3cm} - \fr{2} \log(2+4\alpha+4\alpha^2)(\log(1+x^2)-2 \log(1+2\alpha-x)),\\ 
& \int \frac{\arctan x}{x-1-2\alpha} dx =\ImLiB(x)
-\fr{2} \arctan x (\log(1+x^2)-\log(2+4\alpha+4\alpha^2)
\nonum
& \hspace{3cm}-\fr{2} \arctan(1+2\alpha) (\log(1+x^2)-2\log(1+2\alpha-x)).
\end{align}

\section{Quasi-local stress energy tensor}\label{sec:brownyork}
Here we present the quasi-local stress energy tensor of the EGB black string defined by Brown-York's method
\begin{align}
 {\sf T}^{\mu\nu} := \lim_{r \to \infty} \frac{r^{n+1}}{8\pi G} \left(K h^{\mu\nu}-K^{\mu\nu}\right)-({\rm regulator}), 
\end{align}
where $h_{\mu\nu}$ and $K_{\mu\nu}$ is the metric and extrinsic curvature of a $r$-constant surface. At large $D$, we normalize the tensor so that it remains finite at the limit
\begin{align}
 {\sf T}^{\mu\nu} = \frac{n}{16\pi G} T^{\mu\nu}
\end{align}
where the normalized components up to NLO are given by
\begin{align}
& T^{tt}= (\alpha+1) m
\nonum
&\quad+\fr{n} \left[ \beta  m- \frac{\alpha p^2}{m} + (1+2\alpha-\beta ) \frac{ p \partial_z m}{m} + (\beta -4\alpha-3) \partial_z p-(2\alpha m+(\alpha+1) \partial_z p)\log m \right],\label{eq:defTtt}
\end{align}
\begin{align}
&T^{tz}=(\alpha+1)(p-\partial_z m) \nonum
&\hspace{1cm}+ \fr{n}\left[-\left(\beta +\frac{2\alpha^2-3\alpha-1}{2\alpha+1}\right) p-\frac{\alpha p^3}{m^2} +\left(\frac{10 \alpha ^3+10 \alpha ^2+5 \alpha +1}{(2 \alpha +1) \left(2 \alpha ^2+2 \alpha
   +1\right)}-\beta \right) \frac{p (\partial_z m)^2}{m^2}\right.\nonum
&   \left. \quad - \left(\beta +\frac{\alpha(2\alpha-1)}{2\alpha+1}\right)\frac{  p \partial_z p }{m}+\left(\beta +\frac{4 \alpha ^4-4 \alpha ^3-6 \alpha ^2-4 \alpha
   -1}{(2 \alpha +1) \left(2 \alpha ^2+2 \alpha
   +1\right)}\right) \partial_z m\right.\nonum
& \left. \quad    +\left(\beta+\frac{6\alpha^2-3\alpha-1}{2(2\alpha+1)}\right)\frac{ p^2 \partial_z m }{m^2}+\left(\frac{4 \alpha ^4}{(2 \alpha +1) \left(2 \alpha ^2+2 \alpha +1\right)}+\beta \right)\frac{\partial_z p \partial_z m }{m}\right.\nonum
&\left.\hspace{2cm} +\left(\frac{2\alpha^2+\alpha+1}{1+2\alpha}\partial_z^2 p+(1+\alpha)\frac{ p^2 \partial_z m-2 mp\partial_z p}{m^2}-2\alpha p\right. \right. \nonum
& \left. \left.\hspace{2cm}\qquad+\frac{2\alpha(mp\partial_z^2+m\partial_z p \partial_z m- p (\partial_z m)^2+2\alpha m^2 \partial_z m)}{(2\alpha+1)m^2}  \right)\log m\right],
\end{align}
\begin{align}
&T^{zz}=-\frac{2 \alpha p \partial_z  m }{(2 \alpha 
   +1)m}+(\alpha +1)\partial_t   m+\frac{(\alpha +1)
   p^2}{m}-\frac{\left(2 \alpha ^2+\alpha
   +1\right) (m+\partial_z p)}{2 \alpha +1}\nonum
& + \fr{n}\left[\left(\frac{-4 \alpha ^4+8 \alpha ^3+15 \alpha ^2+6 \alpha +1}{(\alpha +1) (2 \alpha
   +1)^2}-\frac{\left(2 \alpha ^2+\alpha +1\right) \beta }{(2 \alpha+1)(\alpha+1)}\right)m+\left(3 \beta -\frac{2 \alpha ^3+9 \alpha ^2+13 \alpha +4}{2 \alpha ^2+3 \alpha +1}\right)\frac{p^2}{m}\right.\nonum
   &-\frac{\alpha p^4}{m^3}
    + \frac{\gamma_1 p (\partial_z m)^3}{m^3}
    +\frac{\gamma_2(\partial_z p)^2}{m} 
    + \gamma_3\frac{m\partial_z^2 m-2(\partial_z m)^2}{m}
    +\frac{\gamma_4 p^2 (\partial_z m)^2}{m^3}
    +\frac{\gamma_5\partial_z p(\partial_z m)^2}{m^2}\nonum
&   + \frac{\gamma_6 p^2 \partial_z^2 m}{m^2}
+ \frac{\gamma_7 p \partial_z^2 p}{m} 
+ \gamma_8\frac{m \partial_z^3 p-p \partial_z^3 m}{m}  
+ \left(\gamma_9+\frac{4\alpha^2+2\alpha+1}{2\alpha+1}\frac{p^2}{m^2}\right)\partial_t m \nonum
& -\frac{2(4\alpha^2+3\alpha+1)}{2\alpha+1}\frac{p\partial_t p}{m}
+ \gamma_{10} \partial_z p + \left(\frac{2 \alpha ^3+10 \alpha ^2+13 \alpha +3}{(\alpha +1) (2 \alpha +1)}-3 \beta\right) \frac{p^2 \partial_z p}{m^2}+\gamma_{11}\frac{\partial_z p \partial_z^2 m}{m}\nonum
&+\gamma_{12}\frac{\partial_t m \partial_z p}{m}+\gamma_{13} \frac{m\partial_t \partial_z p-p \partial_t \partial_z m- \partial_z m \partial_t p}{m}+ \gamma_{14} \frac{p\partial_z m}{m}+ \left(3 \beta -\frac{7 \alpha ^2+12 \alpha +3}{(\alpha +1) (2 \alpha +1)}\right) \frac{p^3 \partial_z m}{m^3}\nonum
&+ \gamma_{15} \frac{p\partial_z p \partial_z m}{m^2}
+ \gamma_{16} \frac{p\partial_z^2 m \partial_z m}{m^2}
+ \gamma_{17} \frac{\partial_z^2 p \partial_z m }{m}
+ \gamma_{18} \frac{p \partial_z m \partial_t m}{m^2}\nonum
   &\left.+\log m  \left( \frac{2\alpha\partial_z m}{2\alpha+1}  \left(\frac{   p\partial_t m  }{  m ^2}
-\frac{   \partial_t p }{ m }-\frac{  (2 \alpha -1)
   p }{(2 \alpha +1)   m }\right)-\partial_t m  \left(\frac{(\alpha
   +1) p ^2}{m ^2}+\frac{4 \alpha ^2}{2 \alpha
   +1}\right)\right.\right. \nonum
   &\left.\left. \hspace{2cm}-\frac{2 \alpha p \partial_t \partial_z m  }{(2
   \alpha  +1)m }+\frac{2 (\alpha +1)   p \partial_t p   }{m }-\frac{2 \alpha    p ^2}{m }+\frac{2 \alpha(4\alpha^2+4\alpha-1)m}{(2\alpha+1)^2} \right.\right.\nonum
&   \left. \left. \hspace{5cm}+\frac{\left(4 \alpha
   ^3+4 \alpha ^2-5 \alpha -1\right)
    \partial_z p }{(2 \alpha +1)^2}-\frac{\left(2
   \alpha ^2+\alpha +1\right)  \partial_t \partial_z p }{2 \alpha
   +1}\right)\right]   
\end{align}
where
\begin{align}
& \beta = \frac{\log \left(2 \alpha ^2+2 \alpha +1\right)}{4
   \alpha +2}-\log (\alpha +1)+\arctan(2 \alpha
   +1)-\frac{\pi }{4}+1,
\end{align}
and
\begin{align}
&\gamma_1=\frac{4 \alpha  \left(24 \alpha ^4+38 \alpha ^3+26 \alpha ^2+9 \alpha   +1\right)}{(\alpha +1) (2 \alpha +1)^2 \left(2 \alpha ^2+2 \alpha +1\right)}-\frac{16   \alpha  \log (\alpha +1)}{2 \alpha +1},\nonum
&\gamma_2=\frac{\left(2 \alpha ^2+\alpha +1\right) \beta   }{2 \alpha ^2+3 \alpha +1}-\frac{24 \alpha ^6+96 \alpha ^5+138 \alpha ^4+108 \alpha   ^3+53 \alpha ^2+17 \alpha +2}{(\alpha +1) (2 \alpha +1)^2 \left(2 \alpha ^2+2 \alpha   +1\right)}+\frac{4 \alpha  \log (\alpha +1)}{2 \alpha +1},\nonum
&\gamma_3=-\frac{10 \alpha ^3+20 \alpha
   ^2+11 \alpha +2}{4 \alpha ^3+6 \alpha ^2+4 \alpha +1}+\frac{4 \alpha  \log (\alpha
   +1)}{2 \alpha +1}+2 \beta ,\nonum 
&\gamma_4=-\frac{2 \left(2 \alpha ^2+6 \alpha +1\right) \beta }{2
   \alpha ^2+3 \alpha +1}-\frac{2 \alpha  \left(16 \alpha ^5+56 \alpha ^4+44 \alpha ^3+4
   \alpha ^2-10 \alpha -3\right)}{(\alpha +1) (2 \alpha +1)^2 \left(2 \alpha ^2+2 \alpha
   +1\right)}+\frac{8 \alpha  \log (\alpha +1)}{2 \alpha +1},\nonum
& \gamma_5 =   \frac{16 \alpha  \log (\alpha
   +1)}{2 \alpha +1}-\frac{4 \alpha  \left(24 \alpha ^4+38 \alpha ^3+26 \alpha ^2+9 \alpha
   +1\right)}{(\alpha +1) (2 \alpha +1)^2 \left(2 \alpha ^2+2 \alpha +1\right)}
,   \nonum
&\gamma_6 =-\frac{\left(2 \alpha ^2+\alpha +1\right) \beta }{2 \alpha ^2+3 \alpha +1}+\frac{2
   \alpha ^3+13 \alpha ^2+6 \alpha +3}{4 \alpha ^2+6 \alpha +2}-\frac{4 \alpha  \log   (\alpha +1)}{2 \alpha +1},\nonum
&\gamma_7  = \frac{\left(2 \alpha ^2+\alpha +1\right) \beta }{2 \alpha ^2+3
   \alpha +1}-\frac{2 \alpha ^3+9 \alpha ^2+5 \alpha +2}{2 \alpha ^2+3 \alpha +1}+\frac{4   \alpha  \log (\alpha +1)}{2 \alpha +1},\nonum
&\gamma_8 =   -\frac{3 \alpha +1}{\alpha +1}+\frac{4 \alpha 
   \log (\alpha +1)}{2 \alpha +1}+\beta,\nonum
&\gamma_9 =      -\frac{8 \alpha ^6+16 \alpha ^5+42 \alpha ^4+62
   \alpha ^3+41 \alpha ^2+12 \alpha +1}{(\alpha +1) (2 \alpha +1)^2 \left(2 \alpha ^2+2   \alpha +1\right)}+\frac{4 \alpha  \log (\alpha +1)}{2 \alpha +1}+\beta,\nonum
&\gamma_{10} =\frac{4 \alpha 
   \log (\alpha +1)}{2 \alpha +1}-\frac{32 \alpha ^6+88 \alpha ^5+96 \alpha ^4+58 \alpha
   ^3+24 \alpha ^2+9 \alpha +1}{(\alpha +1) (2 \alpha +1)^2 \left(2 \alpha ^2+2 \alpha
   +1\right)},\nonum 
&\gamma_{11} = \frac{26 \alpha ^4+36 \alpha ^3+23 \alpha ^2+8 \alpha +1}{4 \alpha ^4+10 \alpha
   ^3+10 \alpha ^2+5 \alpha +1}-\frac{8 \alpha  \log (\alpha +1)}{2 \alpha +1}-\beta,\nonum
& \gamma_{12} = \frac{\left(4 \alpha ^2+4 \alpha +2\right) \beta }{2 \alpha ^2+3 \alpha +1}-\frac{8
   \alpha ^6+16 \alpha ^5+50 \alpha ^4+82 \alpha ^3+61 \alpha ^2+22 \alpha +3}{(\alpha +1)
   (2 \alpha +1)^2 \left(2 \alpha ^2+2 \alpha +1\right)},\nonum
& \gamma_{13} =\frac{1-2 \alpha +4 \alpha  \log (\alpha +1)}{2 \alpha +1}-\frac{\left(2 \alpha ^2+\alpha
   +1\right) \beta }{2 \alpha ^2+3 \alpha +1},\nonum
 &  \gamma_{14}=5 \beta +\frac{32 \alpha ^6+40 \alpha ^5-68 \alpha ^4-154 \alpha ^3-114 \alpha ^2-37  \alpha -5}{(\alpha +1) (2 \alpha +1)^2 \left(2 \alpha ^2+2 \alpha +1\right)},\nonum
 &\gamma_{15}=\frac{\left(2 \alpha ^2+11 \alpha +1\right) \beta }{2 \alpha ^2+3 \alpha +1}+\frac{64
   \alpha ^6+240 \alpha ^5+280 \alpha ^4+166 \alpha ^3+60 \alpha ^2+19 \alpha +3}{(\alpha
   +1) (2 \alpha +1)^2 \left(2 \alpha ^2+2 \alpha +1\right)}-\frac{12 \alpha  \log (\alpha
   +1)}{2 \alpha +1},\nonum
&\gamma_{16}   =-\frac{120 \alpha ^5+244 \alpha ^4+226 \alpha ^3+114 \alpha ^2+29 \alpha +3}{(\alpha   +1) (2 \alpha +1)^2 \left(2 \alpha ^2+2 \alpha +1\right)}+\frac{20 \alpha  \log (\alpha   +1)}{2 \alpha +1}+3 \beta ,\nonum
&\gamma_{17}=\frac{68 \alpha ^5+146 \alpha ^4+144 \alpha ^3+75 \alpha
   ^2+19 \alpha +2}{(\alpha +1) (2 \alpha +1)^2 \left(2 \alpha ^2+2 \alpha
   +1\right)}-\frac{12 \alpha  \log (\alpha +1)}{2 \alpha +1}-2 \beta,\nonum
&\gamma_{18} =   -\frac{\left(6
   \alpha ^2+5 \alpha +3\right) \beta }{2 \alpha ^2+3 \alpha +1}+\frac{-24 \alpha ^5-20
   \alpha ^4+22 \alpha ^3+34 \alpha ^2+17 \alpha +3}{(\alpha +1) (2 \alpha +1)^2 \left(2
   \alpha ^2+2 \alpha +1\right)}+\frac{4 \alpha  \log (\alpha +1)}{2 \alpha +1}.
\end{align}

\end{document}